\documentclass[onecolumn,a4paper,accepted=2023-01-17]{quantumarticle}
\pdfoutput=1
\usepackage{preamble}
\usepackage{xcolor}
\usepackage[normalem]{ulem}
\usepackage{realboxes}
\definecolor{bkgd}{RGB}{240,242,246}
\definecolor{orange-red}{rgb}{1.0, 0.27, 0.0}

\newcommand{\markedup}[1]{\textcolor{black}{#1}}

\newcommand{\fancylink}[2]{\colorbox{bkgd}{\color{orange-red}\href{#1}{\sf {#2}}}}
\newcommand{\rooturl}{https://github.com/tencent-quantum-lab/tensorcircuit/blob/master/}
\newcommand{\docurl}{https://tensorcircuit.readthedocs.io/en/latest/}

\begin{document}
\title{TensorCircuit: a Quantum Software Framework for the NISQ Era}

\author{Shi-Xin Zhang}
\affiliation{Tencent Quantum Laboratory, Tencent, Shenzhen, Guangdong 518057, China}
%\affiliation{Institute for Advanced Study, Tsinghua University, Beijing 100084, China} 

\author{Jonathan Allcock}
\affiliation{Tencent Quantum Laboratory, Tencent, Hong Kong, China}

\author{Zhou-Quan Wan}
\affiliation{Institute for Advanced Study, Tsinghua University, Beijing 100084, China}
\affiliation{Tencent Quantum Laboratory, Tencent, Shenzhen, Guangdong 518057, China}

\author{Shuo Liu}
\affiliation{Institute for Advanced Study, Tsinghua University, Beijing 100084, China}
\affiliation{Tencent Quantum Laboratory, Tencent, Shenzhen, Guangdong 518057, China}

\author{Jiace Sun}
\affiliation{Division of Chemistry and Chemical Engineering, California Institute of Technology, Pasadena, CA 91125,
USA}

\author{Hao Yu}
\affiliation{Department of Electrical and Computer Engineering, McGill University, Quebec H3A 0E9
, Canada}

\author{Xing-Han Yang}
\affiliation{Shenzhen Middle School, Shenzhen, Guangdong 518025, China}
\affiliation{Tencent Quantum Laboratory, Tencent, Shenzhen, Guangdong 518057, China}

\author{Jiezhong Qiu}
\affiliation{Tencent Quantum Laboratory, Tencent, Shenzhen, Guangdong 518057, China}

\author{Zhaofeng Ye}
\affiliation{Tencent Quantum Laboratory, Tencent, Shenzhen, Guangdong 518057, China}

\author{Yu-Qin Chen}
\affiliation{Tencent Quantum Laboratory, Tencent, Shenzhen, Guangdong 518057, China}

\author{Chee-Kong Lee}
\affiliation{Tencent America, Palo Alto, California 94306, USA}

\author{Yi-Cong Zheng}
\affiliation{Tencent Quantum Laboratory, Tencent, Shenzhen, Guangdong 518057, China}

\author{Shao-Kai Jian}
\affiliation{Department of Physics, Brandeis University, Waltham, Massachusetts 02453, USA}

\author{Hong Yao}
%\email{yaohong@tsinghua.edu.cn}
\affiliation{Institute for Advanced Study, Tsinghua University, Beijing 100084, China}

\author{Chang-Yu Hsieh}
\email{kimhsieh@tencent.com}
\affiliation{Tencent Quantum Laboratory, Tencent, Shenzhen, Guangdong 518057, China}

\author{Shengyu Zhang}
\email{shengyzhang@tencent.com}
\affiliation{Tencent Quantum Laboratory, Tencent, Shenzhen, Guangdong 518057, China}

% \date{accepted=2023-01-17}
%\date{\today}
% \date{01/21/2023}%{\today}

\begin{abstract}
\tc is an open source quantum circuit simulator based on tensor network contraction, designed for speed, flexibility and code efficiency. Written purely in Python, and built on top of industry-standard machine learning frameworks, \tc supports automatic differentiation, just-in-time compilation, vectorized parallelism and hardware acceleration. These features allow \tc to simulate larger and more complex quantum circuits than existing simulators, and are especially suited to variational algorithms based on parameterized quantum circuits. \tc enables orders of magnitude speedup for various quantum simulation tasks compared to other common quantum software, and can simulate up to 600 qubits with moderate circuit depth and low-dimensional connectivity. With its time and space efficiency, flexible and extensible architecture and compact, user-friendly API, \tc has been built to facilitate the design, simulation and analysis of quantum algorithms in the Noisy Intermediate-Scale Quantum (NISQ) era.
\end{abstract}

\maketitle
\tableofcontents

\section{Introduction}

The landscape of open-source and proprietary software for simulating quantum computers \cite{qcbook} has grown in recent years. While features and functionality vary between packages, across the board users now have many high-quality options available for constructing and simulating quantum circuits. However, to deepen our understanding of quantum algorithm performance, researchers increasingly need to simulate larger and more complex quantum circuits, and to optimize quantum circuits that may contain a large number of tunable parameters.  In spite of the promising state of existing quantum software, there remain a number of challenges in running large-scale, complex simulations. 
Here we introduce \tc, a new open source tensor network based quantum circuit simulator built to address these challenges.  Written in Python, and designed for speed, flexibility and ease-of-use, \tc is built on top of a number of industry leading libraries.  Via the TensorFlow~\cite{abadi2016tensorflow}, JAX~\cite{jax2018github} and PyTorch~\cite{pytorch} machine learning libraries, convenient access is provided for automatic differentiation, just-in-time compilation, vectorized parallelism, and hardware acceleration. Fast tensor network contraction is enabled by the state-of-the-art cotengra~\cite{gray2020, gray2021hyper} package, which also gives users customizable control over the tensor network contraction process. These features enable efficient optimization of parameterized quantum circuits, allowing for more complex cases to be modelled.   In addition, the \tc syntax aims to allow complicated tasks to be implemented with a minimal amount of code, saving time spent coding as well as in simulation.

\subsection{Challenges in simulating quantum circuits}
As fully fault-tolerant quantum computers capable of running large scale quantum algorithms may still be many years away, considerable research effort has been spent investigating the prospects for quantum advantage in the nearer term. Some algorithms for Noisy Intermediate-Scale Quantum (NISQ)~\cite{preskill2018quantum} quantum computers aim to leverage classical computational power to supplement quantum computers, which may have only a limited number of error-prone qubits under control.  In particular, hybrid quantum-classical algorithms \cite{Bharti2021, Cerezo2020b} such as the Variational Quantum Eigensolver (VQE)~\cite{peruzzo2014variational} and the Quantum Approximate Optimization Algorithm (QAOA)~\cite{farhi2014quantum} are based around the concept of parameterized quantum circuits (PQC).  These circuits, which contain quantum gates with tunable parameters (for instance, single-qubit gates with variable rotation angles), are embedded in a classical optimization loop. By optimizing the value of the parameters in the circuit, one aims to drive the output state of the quantum circuit towards the solution to a given problem.  

In a prototypical VQE example, one wishes to find the ground state energy $E_0$ of a quantum system with Hamiltonian $H$. The output of a PQC is a quantum state $\ket{\psi(\boldsymbol{\theta})}$, where $\boldsymbol{\theta}$ is a vector of tunable parameters.  This trial state -- known as an ansatz -- forms a guess for the ground state wavefunction. By performing appropriate measurements on $\ket{\psi(\boldsymbol{\theta})}$, one can estimate the expected energy $\langle H \rangle_{\boldsymbol{\theta}} = \bra{\psi(\boldsymbol{\theta})} H \ket{\psi(\boldsymbol{\theta})}$.  
By minimizing $\langle H \rangle_{\boldsymbol{\theta}}$ with respect to the parameters $\boldsymbol{\theta}$, one obtains an upper bound estimate of the ground state energy:
\eql{
E_0 \le \min_{\boldsymbol{\theta}} \langle H \rangle_{\boldsymbol{\theta}}. \label{eq:variational} \quad 
}

There are a number of issues that affect the efficacy and efficiency for this approach.  Firstly, for an accurate estimate of $E_0$, the ansatz $\ket{\psi(\boldsymbol{\theta})}$ should, for some values of $\boldsymbol{\theta}$, be a good approximation to the true ground state.  Whether this is so depends on the nature of the problem and the complexity of the PQC from which the ansatz is constructed.  On the one hand, simple ansätze  -- for instance, the ``hardware efficient" ansatz of~\cite{kandala2017hardware} -- may be easier to implement on real or simulated quantum computers, but may either not be sufficiently accurate  (for instance, due to lack of physically-relevant structure or short depth) or else suffer from other issues, such as so-called barren plateaus in parameter space~\cite{McClean2018} or local minima in the energy landscape \cite{2109.06957}. On the other hand, more complex ansätze
, such as the unitary coupled cluster (UCC)~\cite{yung2014transistor} approach proposed for quantum chemistry problems, may require quantum circuit complexities and depths beyond what can currently be implemented or simulated, and the associated circuits must then be truncated or simplified.  The ability to simulate larger and deeper quantum circuits would enable a systematic investigation of larger classes of ansätze.

Secondly, for a given ansatz, the evaluation of $\langle H \rangle_{\boldsymbol{\theta}}$ can be an involved process.  For instance, consider the case where $H$ is an $n$-qubit Hamiltonian, which can be expressed as a weighted sum of tensor products of Pauli operators, i.e.,
\eql{
H = \sum_{j=1}^K \alpha_j P_j, \label{eq:H-pauli}
}
where $\alpha_j$ are real coefficients, each Pauli string $P_j$ is of the form $P_j = \sigma_{i_1}\otimes \sigma_{i_2}\otimes\ldots \sigma_{i_n}$, and $\sigma_{i_j}$ are single-qubit Pauli operators or the identity.  In the most straightforward approach, estimating $\langle H \rangle_{\boldsymbol{\theta}}$ is performed by first estimating the expectation of each term $\langle P_j \rangle_{\boldsymbol{\theta}}$ and then adding together these individual contributions.  If the Hamilonian consists of many Pauli terms, ways of speeding up the evaluation of Eq.~\eqref{eq:H-pauli} -- for instance, by exploiting efficient representations of $H$ (e.g. as a sparse matrix or Matrix Product Operator (MPO)~\cite{dmrgmps}), or being able to compute multiple terms in parallel -- can have a large impact on the computation time.

Thirdly, the optimization problem in Eq.~\eqref{eq:variational} is, in general, non-convex.  Thus, finding a global minimum is in general computationally intractable.  However, if gradients of potentially complicated expressions can be efficiently evaluated, one can use 
gradient descent methods to find a local minimum which may yield a decent approximate solution.  By initializing the optimization from a number of different positions (i.e. different values of $\boldsymbol{\theta}$), multiple local minima can be obtained, improving the chances that one of these gives a good solution.  If these multiple solutions can be optimized in parallel, one may achieve large efficiency gains.

\subsection{Machine learning libraries}
The challenges discussed in the previous section overlap to a large degree with problems faced in machine learning, and especially deep learning \cite{Lecun2015}. Fortunately, the need to tackle machine learning problems of increasing complexity and to deal with datasets of ever-larger size, has led to the development of impressive software, and many frameworks are now available that combine powerful features with easy-to-use syntax.  In particular:

\vspace{\baselineskip}\noindent\textbf{Fast gradients.} At the heart of all advanced machine learning packages is the ability to perform automatic differentiation (AD) \cite{Bartholomew-Biggs2000a, GunesBaydin2018}, of which the backpropagation algorithm used to train neural networks is a special case. 
AD enables efficient computation of the gradients of functions defined in code, and is vital to the optimization of many machine learning models.  

\vspace{\baselineskip}\noindent\textbf{JIT.} Just-in-time compilation is a way of compiling certain parts of code during program execution. For interpreted languages such as Python, "jitting" a function can lead to large performance gains, with the time required to execute a jitted function often only a small fraction of the time needed if the function were interpreted. While the first time the function is called there may be an overhead cost required for compilation (staging time), this cost can be negligible to the time saved if the function is subsequently called many times. 

\vspace{\baselineskip}\noindent\textbf{Vectorization (VMAP).} This feature allows a function to be evaluated on multiple points in parallel, with significant speedup compared with using a na\"ive {\sf for} loop. In machine learning this allows one, for instance,  to perform computations on batches of data at the same time. 

\vspace{\baselineskip}\noindent\textbf{Hardware acceleration.} For complex machine learning models, the ability to execute code on multiple CPUs, GPUs and TPUs may be necessary for training to complete in a reasonable amount of time.

\vspace{\baselineskip}These powerful features are also beneficial in simulating quantum computers, and are particularly suited to variational quantum algorithms.  Here, fast gradient evaluation via automatic differentiation gives simulators an inherent advantage over real quantum computers, which must estimate gradients in a less direct way; for instance by sampling the outputs for various input parameter choices and computing finite differences \cite{PhysRevLett.118.150503, PhysRevA.99.032331}.  Vectorization can (among other things) be used to evaluate PQC circuits with multiple parameter choices concurrently, or compute expectations of multiple Pauli strings simultaneously, and JIT and hardware acceleration provide for further time savings.

In addition, there is increasing interest in solving machine learning problems via quantum computing, as well as combining classical and quantum machine learning (QML) algorithms.  Both of these can be facilitated by better integration between classical ML frameworks and quantum circuit simulators, and great value can be derived from a seamless integration of the two.

\subsection{The next phase of quantum software}
While quantum software has progressed a great deal in the last few years -- with packages such as Qiskit~\cite{qiskit}, Cirq~\cite{cirq}, \markedup{ProjectQ~\cite{Steiger2018}}, HiQ~\cite{hiq}, Q\#~\cite{svore2018q}, \markedup{Qibo~\cite{Efthymiou2022}}, and qulacs~\cite{qulacs} all offering powerful functionality and features -- there remain significant advantages to be gained from efficient quantum simulation software supplemented with the power and features of state-of-the-art machine learning.  
Recently, a new generation of quantum software has started to emerge, with TensorFlow Quantum~\cite{tfq}, Pennylane~\cite{pennylane}, Paddle Quantum~\cite{Paddlequantum} and MindQuantum~\cite{mindspore} in the Python ecosystem making inroads here, and providing these features to varying degrees. However, to date, none of these fully combine all of the key features of the previous section with fast quantum circuit simulation. \markedup{That is, there is no good solution for space and time efficient noiseless and noisy quantum simulation that combines a tensor network engine with the machine learning paradigms of AD, JIT and VMAP, as well as GPU support. (See Table~\ref{tab:softwarecomparison} for comparison of quantum software in terms of machine learning paradigm compatibility.)} \tc has been designed to fill this gap, and give users a faster, more flexible and more convenient way to simulate quantum circuits and quantum processes.

\begin{table}\centering
	\begin{tabular}{c|c|c|c|c|c}
		& AD & JIT & VMAP & GPU  & TN\\ \hline
		Qiskit/Cirq/ProjectQ/Qulacs~& $\cdot$  & $\cdot$  &$\cdot$   & \checkmark\checkmark & $\cdot$  \\
%		Qulacs &\checkmark\checkmark  & $\cdot$  & $\cdot$  & \checkmark\checkmark\\
			Qibo & \checkmark & \checkmark\checkmark & $\cdot$  & \checkmark\checkmark & $\cdot$ \\
		TensorFlow Quantum~ & \checkmark & $\cdot$  &  \checkmark\checkmark & $\cdot$ & $\cdot$\\
		Pennylane & \checkmark\checkmark &\checkmark  &  \checkmark& \checkmark\checkmark & $\cdot$ \\
	  {\bf TensorCircuit}   & \checkmark\checkmark &  \checkmark\checkmark&  \checkmark\checkmark &  \checkmark\checkmark  & \checkmark\checkmark\\
	\end{tabular}
	\caption{\markedup{Quantum software support for main machine learning paradigms and tensor network engine. Check marks qualitatively indicate the level of support for the corresponding features: $\checkmark\checkmark $ = good support, $\checkmark$ = limited support, $\cdot$ = not supported. TN is for tensor network simulation engine support. Many well-known quantum software frameworks such as Qiskit or Cirq have no support for AD, JIT and VMAP; Qibo does not support AD with its most optimized and efficient backend; TensorFlow Quantum supports AD of expectation values but not directly of the wavefunction, and does not support quantum circuit simulation on GPU; Pennylane does not support VMAP of trainable parameters and JIT support is fragile as some methods are not JIT-compatible, etc.}}\label{tab:softwarecomparison}
\end{table}

\section{\tc Overview}

Our goal with \tc is to provide the first efficient, tensor network based quantum simulator that is fully compatible with the key features of modern machine learning frameworks, especially the programming paradigms of automatic differentiation, vectorized parallelism and just-in-time compilation. These features are provided via a number of popular machine learning backends which, at the time of writing, are TensorFlow, JAX and PyTorch (see Table.~\ref{fig:backends}).

\begin{table}\centering
\begin{tabular}{c|c|c|c|c}
     & AD & JIT & VMAP & GPU \\ \hline
  TensorFlow & \checkmark\checkmark  & \checkmark\checkmark   & \checkmark\checkmark   & \checkmark\checkmark  \\
  JAX & \checkmark\checkmark   & \checkmark\checkmark   & \checkmark\checkmark   & \checkmark\checkmark  \\
  PyTorch &\checkmark\checkmark  & \checkmark  & \checkmark  & \checkmark\checkmark  \\
NumPy   & $\cdot$  & $\cdot$  & $\cdot$ & $\cdot$   \\
\end{tabular}
\caption{Backends supported by \tc. Check marks qualitatively indicate the level of support for the corresponding features: \markedup{$\checkmark\checkmark$ = good support, $\checkmark$ = limited support,$\cdot$ = not supported.} TensorFlow and JAX both offer comprehensive support for AD, JIT, VMAP and hardware acceleration, and are recommended for most tasks. \markedup{PyTorch currently has limited support for JIT and VMAP as these are implemented in experimental modules and are not yet stable. In \tc we supplement the AD infrastructure of TensorFlow with a VMAP-compatible implementation of Jacobian and Hessian calculations, and enrich the functionality of TensorFlow's VMAP, which in \tc now supports VMAP over multiple arguments. These improvements can be accessed via the unified backend abstraction provided by \tc. }
If no backend is chosen, \tc defaults to using NumPy as a backend, which does not support these advanced ML features. For more on the choice of ML backend, please see the documentation: \fancylink{\docurl faq.html\#which-ml-framework-backend-should-i-use}{faq.html\#which-ml-framework-backend-should-i-use}.}\label{fig:backends}
\end{table}

\begin{figure}[htp]
    \centering
    \fbox{\includegraphics[width=0.8\textwidth]{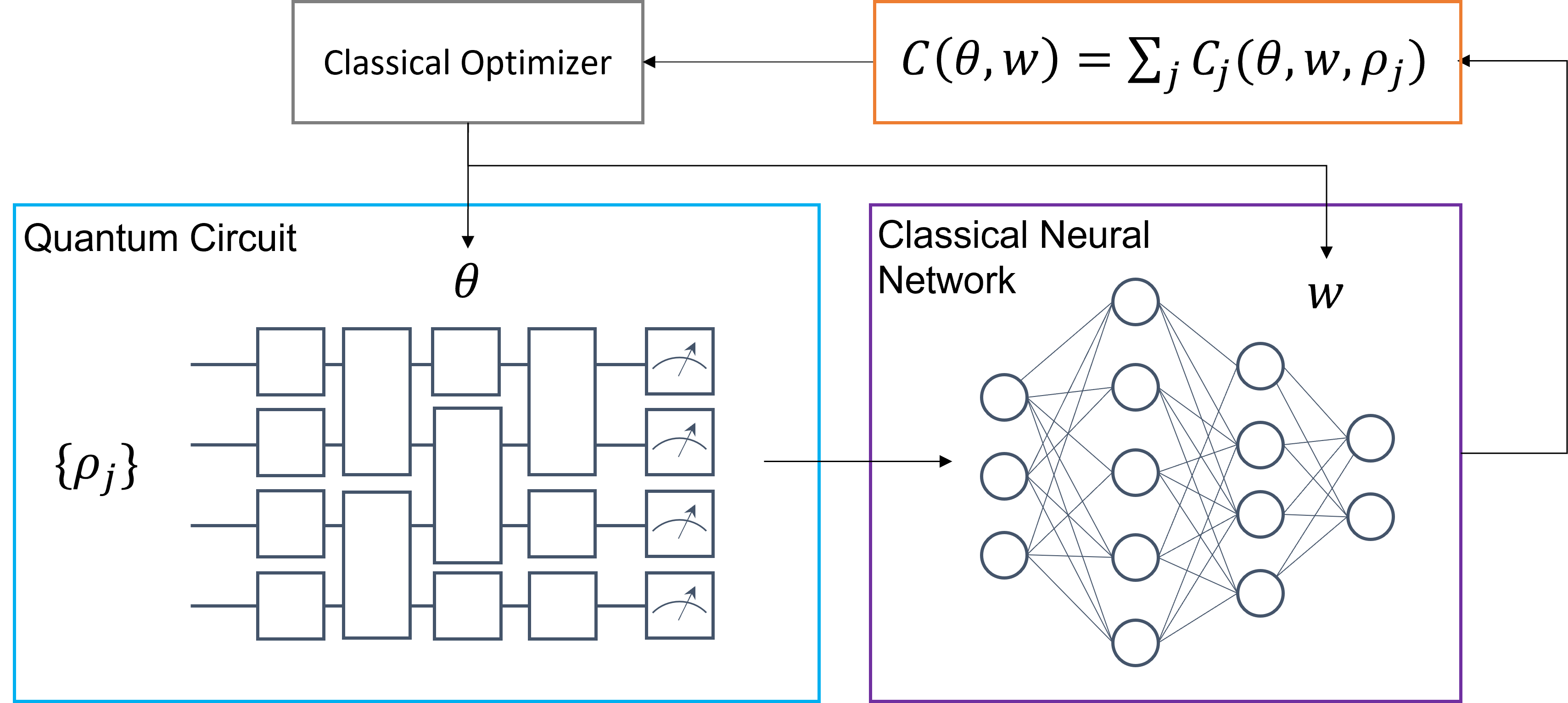}}
    \caption{A general hybrid quantum-classical neural network, where a cost function $C$ is summed over a batch of input states $\{\rho_j\}$ and is dependent on the parameters of a quantum circuit and classical neural network.  By using a classical optimizer, the parameters of both networks can be iteratively improved. Integration with classical machine learning backends allows the entire end-to-end process to be seamlessly simulated in \tc, with {\sf vmap}, {\sf jit} and automatic differentiation enabling efficiency gains throughout the optimization process.}
    \label{fig:architecture}
\end{figure}

Integration with these backends allows for general hybrid quantum-classical models, where the outputs of a parameterized quantum circuit may be fed into a classical neural network or vice versa, and simulated seamlessly (see Figure~\ref{fig:architecture}). This integration is also key for research related to quantum machine learning~\cite{Biamonte2017a}.

Currently, there are very limited options for tensor network based quantum simulators, with most popular software making use of state vector simulators.  State vector simulators are strongly limited by memory as wavefunction amplitudes are stored in full, and can thus struggle to simulate circuits with larger numbers of qubits. On the other hand, quantum circuits with large numbers of (possibly noisy) qubits but relatively short circuit depths -- such as those corresponding to NISQ devices, whose qubits have short coherence times -- fall into the applicable region of tensor network simulators.

\subsection{Design philosophy}

With seamless integration with modern machine learning paradigms supported by an efficient tensor network based simulation engine, the design of \tc has, from the start, been based on the following principles.

 \bp{Speed.} \tc uses tensor network contraction to simulate quantum circuits, and is compatible with state-of-the-art third party contraction engines. In contrast, the majority of current popular quantum simulators are state-vector based.  The tensor contraction framework allows \tc, in many cases, to simulate quantum circuits with improved efficiency in terms of time and space compared with other simulators. JIT, AD, VMAP and GPU support can all also provide significant acceleration (often, of several orders of magnitude) in many scenarios.
    
\bp{Flexibility.}  \tc is designed to be backend agnostic, making it easy to switch between any of the machine learning backends with no change in syntax or functionality. Via different ML backends, there is flexibility to simulate hybrid quantum-classical neural networks, run code on CPUs, GPUs and TPUs, and switch between 32 bit and 64 bit precision data types. 

\bp{Code efficiency.}  Modern machine learning frameworks such as TensorFlow, PyTorch and JAX have user-friendly syntax, allowing for powerful tasks to be carried out with a minimal amount of code. With \tc, we are similarly focused on a compact and easy-to-use API to boost readability and productivity.  Compared with other popular quantum simulation software, \tc can often perform similar tasks with significantly less code (see \fancylink{https://gist.github.com/refraction-ray/754bde72825a1d20ab1bc8dc22d5885e}{tc vs. tfq for VQE} for an example).
The backend agnostic syntax of \tc additionally makes it easy to switch between ML frameworks with a single line of setup code.

\bp{Community focused.}  \tc is open source software, and we care about readability, maintainability and extensibility of the codebase. We invite all members of the quantum computing community to take part in its continued development and use.

\subsection{Tensor network engine}

\tc simulates quantum circuits using the tensor network formalism, which has a long history in computational physics and was more recently pioneered for quantum circuit simulation~\cite{markov2008simulating} . Indeed, the graphical representations for quantum circuits and tensor networks are consistent, rendering a direct and simple translation from quantum circuit simulation to tensor network contraction.
In this picture, quantum circuits are represented by a network of low-rank tensors corresponding to individual quantum gates, and the computation of amplitudes, expectation values or other scalar quantities is performed by contracting the edges of the network until only a single node remains.  The order, or path, in which the edges are contracted is important, and can have a large impact on the time and space required to contract the network \cite{gray2021hyper, 2110.09894, 2204.09052}. See~\cite{orus2014practical} for a good introduction to tensor networks from a physics perspective.

In the most general case, as for other types of quantum simulators, the time required to simulate quantum circuits via tensor network contraction is exponential in the number of qubits. However, in special cases -- including many of practical relevance -- tensor network contraction can offer significant advantages since it avoids the memory bottleneck that plagues full state simulators and, to date, the largest scale quantum computing simulations such as the simulation of random circuits used in quantum supremacy experiments~\cite{Arute2019, PhysRevLett.127.180501} have all been performed via this approach \cite{PhysRevLett.123.190501, PhysRevLett.128.030501, Liu_2021, 2005.06787,  2111.01066, 2111.03011}.

\begin{figure}[htp]
    \centering
\includegraphics[width=0.8\textwidth]{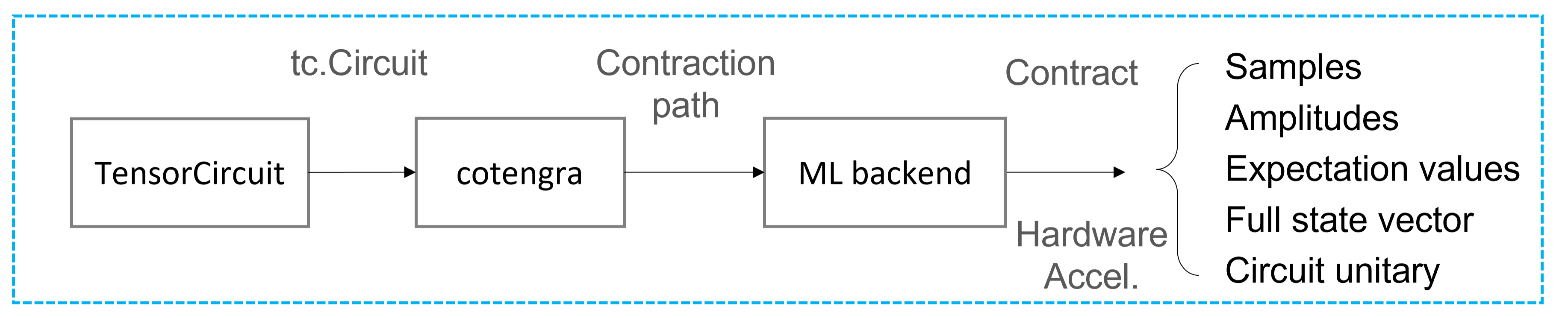}
    \caption{Schematic of the quantum circuit components in Figure~\ref{fig:architecture}. In \tc, the gates that comprise the quantum circuit are contained in a {\sf tc.Circuit} object. Computation of circuit outputs is executed in two steps. First, a tensor contraction path is determined by a contraction engine, e.g., cotengra, and the backends take care of the actual contraction.}
    \label{fig:workflow}
\end{figure}

The tensor network data structure used in \tc is powered by the TensorNetwork~\cite{tensornetwork} package. In addition, \tc can utilize state-of-the-art external Python packages such as cotengra for selecting efficient contraction paths, and the contraction is then performed via {\sf einsum} and {\sf matmul} by the machine learning backend selected by the user (see Figure~\ref{fig:workflow}).

\begin{figure}[htp]
    \centering
\includegraphics[width=0.9\textwidth]{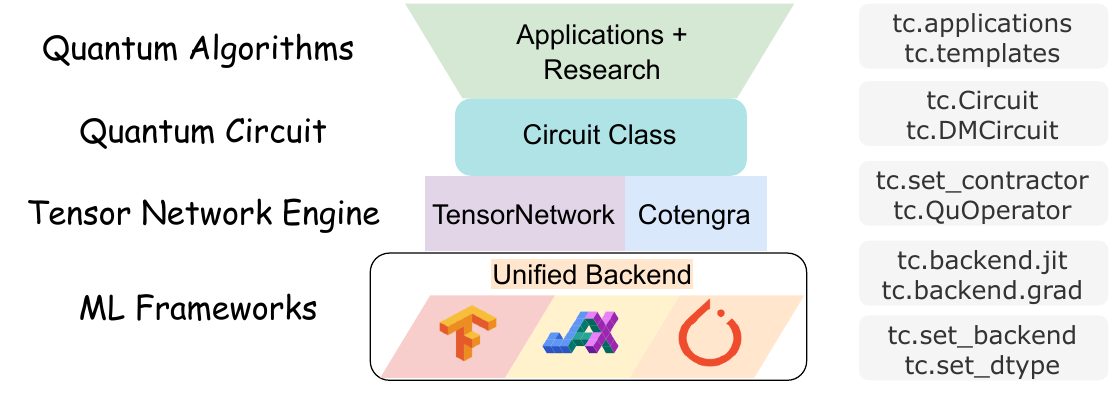}
    \caption{Software architecture of \tc. The abstraction layers are shown on the left while representative APIs are shown on the right. At the bottom, different machine learning frameworks such as TensorFlow, JAX and PyTorch are unified via a set of backend agnostic APIs. These backends, together with the tensor network engine, enable efficient quantum circuit simulation and implementation of quantum-classical hybrid algorithms and applications.  }
    \label{fig:software}
\end{figure}

{\bp {Architecture}}. The tensor network engine underlying the simulation of quantum circuits is built on top of various machine learning frameworks with an abstraction layer in between that unifies different backends. At the application layer, \tc also includes various advanced quantum algorithms based on our latest research \cite{arxiv.2010.08561, Zhang2021_np, Zhang2021_prl, 2112.10380, 2111.13719, arXiv:2208.02866, arXiv:2210.09007}. The overall software architecture is shown in Figure~\ref{fig:software}.

\subsection{Installing and contributing to \tc}
 \tc is open-sourced under the Apache 2.0 license. The software is available on the Python Package Index (PyPI) and can be installed using the command {\sf pip install tensorcircuit}.

The development of \tc is open-sourced and centered on GitHub: \fancylink{https://github.com/tencent-quantum-lab/tensorcircuit}{\tc Repository}.  We welcome all members of the quantum community to contribute, whether it is 
\begin{itemize}
\item Answering questions on the discussion page or issues page.
\item Raising issues such as bug reports or feature requests on the issues page.
\item Improving the documentation (docstrings/tutorials) by pull requests.
\item Contributing to the codebase by pull requests.
\end{itemize}

For more details, please refer to the contribution guide: \fancylink{\docurl contribution.html}{contribution.html}. 

\section{Circuits and gates} 

\begin{mdframed}
\textbf{Jupyter notebook: }
\fancylink{\rooturl docs/source/whitepaper/3-circuits-gates.ipynb}{3-circuits-gates.ipynb}
\end{mdframed}

In \tc, a quantum circuit on $n$ qubits -- which supports both noiseless and noisy simulations via Monte Carlo trajectory methods --  is created by the {\sf tc.Circuit(n)} API. Here we show how to create basic circuits, apply gates to them, and compute various outputs.  

\subsection{Preliminaries}

In the remainder of this document, we assume that we have both \tc and NumPy imported as
\begin{lstlisting}[language=Python]
import tensorcircuit as tc
import numpy as np
\end{lstlisting}
Furthermore we assume that a \tc backend has been set, e.g.
\begin{lstlisting}[language=Python]
K = tc.set_backend("tensorflow")
\end{lstlisting}
and the symbol {\sf K} that appears in code snippets (e.g. {\sf K.real()}) refers to that backend. Other options for the {\sf set\_backend} method are ``jax'', ``pytorch'' and "numpy" (the default backend).

In \tc, qubits are numbered from $0$, with multiqubit registers numbered with the zeroth qubit on the left, e.g. $\ket{0}_{q0}\ket{1}_{q1}\ket{0}_{q2}$. Unless needed, we will omit subscripts and use compact notation e.g. $\ket{010}$ to denote multiqubit states. $X,Y,Z$ denote the standard single qubit Pauli operators, with subscripts e.g. $X_3$ to clarify which qubit is acted on. Expectation values of operators with respect to a state $\ket{\psi}$ such as $\bra{\psi} Z\otimes I \otimes X\ket{\psi}$ will be denoted in shorthand as  $\langle Z_0 X_2\rangle$. Unless stated, expectation values are always with respect to the output state of a given quantum circuit. If that circuit is parameterized by a set of angles $\boldsymbol{\theta}$, then the parameter-dependent expectation value may be denoted by $\langle \cdot \rangle_{\boldsymbol{\theta}}$.

In \tc the default runtime datatype is complex64, but if higher precision is required this can be set as follows:
\begin{lstlisting}[language=Python]
tc.set_dtype("complex128")
\end{lstlisting}

\subsection{Basic circuits and outputs}\label{sec:basics}
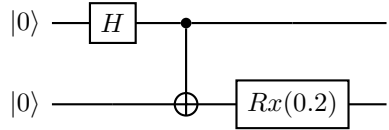
\begin{figure}[h!]
\centering
\begin{quantikz}
\lstick{$\ket{0}$}&\gate{H} &\ctrl{1} &\qw &\qw \\
\lstick{$\ket{0}$}&\qw &\targ{} &\gate{Rx(0.2)} &\qw 
\end{quantikz}\caption{A two-qubit circuit consisting of Hadamard, CNOT and single-qubit $R_X$ rotation.} \label{fig:simple-circuit}
\end{figure}

 Consider the following two-qubit quantum circuit consisting of a Hadamard gate on qubit q0, a CNOT on qubits q0 and q1, and a single qubit rotation $R_X(0.2)$ of qubit q1 by angle $0.2$ about the $X$ axis (see Figure~\ref{fig:simple-circuit}). 
Qubits are numbered from $0$ with $q_0$ on the top row. 
This circuit can be implemented in \tc as
\begin{lstlisting}[language=Python]
c = tc.Circuit(2)
c.h(0)
c.cnot(0, 1)
c.rx(1, theta=0.2)
\end{lstlisting}
From this, various outputs can be computed.

\bp{Basic outputs.}
The full wavefunction can be obtained via
\begin{lstlisting}[language=Python]
c.state()
\end{lstlisting}
which will output an array $[\alpha_{00}, \alpha_{01}, \alpha_{10},\alpha_{11}]$ corresponding to the amplitudes of the $\ket{00}, \ket{01},\ket{10},\ket{11}$ basis states.  The full wavefunction can also be used to generate the reduced density matrix of a subset of the qubits, e.g.
\begin{lstlisting}[language=Python]
# reduced density matrix for qubit 1
s = c.state()
tc.quantum.reduced_density_matrix(s, cut=[0])  # cut: list of qubit indices to trace out
\end{lstlisting}
Amplitudes of individual basis vectors are computed by passing the corresponding bit-string value to the {\sf amplitude} function. For example, the amplitude of the $\ket{10}$ basis vector is computed by
\begin{lstlisting}[language=Python]
c.amplitude("10")
\end{lstlisting}
The unitary matrix corresponding to the entire quantum circuit can also be output: 
\begin{lstlisting}[language=Python]
c.matrix()
\end{lstlisting}

\bp{Measurements and samples.}
Random samples corresponding to $Z$-measurements , i.e. in the $\{\ket{0}, \ket{1}\}$ basis, on all qubits can be generated using
\begin{lstlisting}[language=Python]
c.sample()
\end{lstlisting}
which will output a $(\text{bitstring}, \text{probability})$ tuple, comprising a binary string corresponding to the measurement outcomes of a Z measurement on all the qubits and the associated probability of obtaining that outcome.  Z measurements on a subset of qubits can be performed with the {\sf measure} command
\begin{lstlisting}[language=Python]
# return (outcome, probability) of measuring qubit 0 in Z basis
print(c.measure(0, with_prob=True))
\end{lstlisting}
For measurement of multiple qubits, simply provide a list of indices to {\sf measure}, e.g. if {\sf c} were a $4$-qubit circuit, measurement of qubits $1,3$ can be done via
\begin{lstlisting}[language=Python]
c.measure(1, 3, with_prob=True)
\end{lstlisting}
Note that measurement gates do not need to be explicitly added to the circuit in order to compute these outcomes and the {\sf measure} and {\sf sample} commands do not collapse the circuit output state. Measurement gates can be added and used, for instance, when gates must be applied conditioned on mid-circuit measurement outcomes (see Section~\ref{sec:cond-measure}).

\vspace{\baselineskip}\noindent\textbf{Expectation values.} Expectation values such as $\langle X_0 \rangle$, $\langle X_1 + Z_1\rangle$ or $\langle Z_0 Z_1\rangle$ can be computed via the ${\sf expectation}$ method 
of a circuit object, where the operator is defined via {\sf Gate} object or simply an array. 
\begin{lstlisting}[language=Python]
c.expectation([tc.gates.x(), [0]])                       # <X0>
c.expectation([tc.gates.x() + tc.gates.z(), [1]])        # <X1 + Z1>
c.expectation([tc.gates.z(), [0]], [tc.gates.z(), [1]])  # <Z0 Z1>
\end{lstlisting}
and expectations of user-defined operators can also be computed by supplying the corresponding array of matrix elements. For instance, the operator $2X + 3Z$ can be expressed as a matrix as
\eq{
\begin{pmatrix}
3 & 2 \\
2 & -3
\end{pmatrix}
}
and implemented (assuming the observable is measured on qubit 0) as
\begin{lstlisting}[language=Python]
c.expectation([np.array([[3, 2], [2, -3]]), [0]])
\end{lstlisting}

\vspace{\baselineskip}\noindent\textbf{Expectations of Pauli strings.}
While expectations of products of Pauli operators, e.g. $\langle Z_0 X_1\rangle$ can be computed using {\sf c.expectation} as above, \tc provides another way of computing such expressions which may be more convenient for longer Pauli strings: 
\begin{lstlisting}[language=Python]
c.expectation_ps(x=[1], z=[0])
\end{lstlisting}
and longer Pauli strings can similarly be computed by providing lists of indices corresponding to the qubits that the $X,Y,Z$ operators act on. For example, for an $n=5$ qubit circuit, the expectation value $\langle Z_0 X_1 Z_2 Y_4\rangle $ %= \bra{\psi} z \otimes z \otimes z \otimes I \otimes y\ket{\psi}
is computed as
\begin{lstlisting}[language=Python]
c.expectation_ps(x=[1], y=[4], z=[0, 2])
\end{lstlisting}

\bp{Standard quantum gates.}
Beyond the CNOT, Hadamard and $R_X$ gates we have encountered so far, \tc provides support for a wide variety of commonly encountered quantum gates.  The full list of gates can be found by querying
\begin{lstlisting}[language=Python]
tc.Circuit.sgates # non-parameterized gates
tc.Circuit.vgates # parameterized gates
\end{lstlisting}
The matrix corresponding to a given gate, e.g. the Hadamard {\sf h} gate, can be accessed in the following way
\begin{lstlisting}[language=Python]
tc.gates.matrix_for_gate(tc.gates.h())
\end{lstlisting}

\bp{Arbitrary unitaries.} User-defined unitary gates may be implemented by specifying their matrix elements as an array. As an example, the unitary $S = \begin{pmatrix} 1 & 0 \\  0 & i\end{pmatrix}$ -- which can also directly be added by calling {\sf c.s()} --
can be implemented as
\begin{lstlisting}[language=Python]
c.unitary(0, unitary = np.array([[1,0],[0,1j]]), name='S')
\end{lstlisting}
where the optional {\sf name} argument specifies how this gate is displayed when the circuit is output to \LaTeX.

\bp{Exponential gates.} Gates of the form $e^{i\theta G}$ where matrix $G$ satisfies $G^2 = I$ admit a fast implementation via the {\sf exp1} command, e.g., the two qubit gate $e^{i\theta Z\otimes Z}$ acting on qubits $0$ and $1$
\begin{lstlisting}[language=Python]
c.exp1(0, 1, theta=0.2, unitary=tc.gates._zz_matrix)
\end{lstlisting}
where {\sf tc.gates.\_zz\_matrix} creates a numpy array corresponding to the the matrix
\eq{
Z\otimes Z = \begin{pmatrix}
1 & 0 & 0 & 0 \\
0 & -1 & 0 &  0 \\
0 & 0 & -1 & 0 \\
0 & 0 & 0 & 1
\end{pmatrix}
.}
General exponential gates, where $G^2\neq I$ can be implemented via the {\sf exp} command:
\begin{lstlisting}[language=Python]
c.exp(0, theta=0.2, unitary=np.array([[2, 0],[0, 1]]))
\end{lstlisting}

\bp{Non-unitary gates.}
\tc also supports the application of non-unitary gates by supplying a non-unitary matrix as the argument to {\sf c.unitary}, e.g.
\begin{lstlisting}[language=Python]
c.unitary(0, unitary=np.array([[1,2],[2,3]]), name='non_unitary')
\end{lstlisting}

Note that non-unitary gates will lead to an output state that is no longer normalized, since normalization is often unnecessary and takes additional computational time.

\subsection{Specifying the input state and composing circuits}

By default, quantum circuits are applied to the initial all-zero product state.  Arbitrary initial states can be set by passing an array containing the input state amplitudes to the optional {\sf inputs} argument of {\sf tc.Circuit}.  For example, the  maximally entangled state $\frac{\ket{00}+\ket{11}}{\sqrt{2}}$ can be input as follows:
\begin{lstlisting}[language=Python]
c1 = tc.Circuit(2, inputs=np.array([1, 0, 0, 1] / np.sqrt(2)))
\end{lstlisting}
Input states in Matrix Product State (MPS) form can also be input via the optional {\sf mps\_inputs} argument of {\sf tc.Circuit}. See Section~\ref{sub:mps} for details.

Circuits that act on the same number of qubits can be composed together via the {\sf c.append()} or {\sf c.prepend()} commands. With {\sf c1} defined as above, we can create a new circuit {\sf c2} and then compose them together:
\begin{lstlisting}[language=Python]
c2 = tc.Circuit(2)
c2.cnot(0, 1)

c3 = c1.append(c2)
\end{lstlisting}
This leads to a circuit $C_3$ which is equivalent to first applying $C_1$ and then $C_2$.

\subsection{Circuit transformation and visualization}
{\sf tc.Circuit} objects can be converted to and from Qiskit {\sf QuantumCircuit} objects.  Export to Qiskit is done by
\begin{lstlisting}[language=Python]
c.to_qiskit()
\end{lstlisting}
and the resulting {\sf QuantumCircuit} object can then be compiled and run on compatible physical quantum processors and simulators.  Conversely, importing a {\sf QuantumCircuit} object from Qiskit is done via
\begin{lstlisting}[language=Python]
c = tc.Circuit.from_qiskit(QuantumCircuit)
\end{lstlisting}

There are two ways to visualize quantum circuits generated in \tc.  The first is to use
\begin{lstlisting}[language=Python]
print(c.tex())
\end{lstlisting}
which outputs the code for drawing the associated quantum circuit using the  \LaTeX  quantikz package~\cite{kay2018tutorial}. The second method uses the draw function:
\begin{lstlisting}[language=Python]
c.draw()
\end{lstlisting}
which is a shortcut for
\begin{lstlisting}[language=Python]
qc = c.to_qiskit()
qc.draw()
\end{lstlisting}

Underlying circuit transformation and visualization utilities is the quantum intermediate representation (IR) for \tc {\sf Circuit} objects, which can be obtained by
\begin{lstlisting}[language=python]
c.to_qir()
\end{lstlisting}

\section{Gradients, optimization and variational algorithms}

\begin{mdframed}
\textbf{Jupyter notebook: }
\fancylink{\rooturl docs/source/whitepaper/4-gradient-optimization.ipynb}{4-gradient-optimization.ipynb}
\end{mdframed}

\tc is designed to make optimization of parameterized quantum gates easy, fast and convenient.  Consider a variational circuit acting on $n$ qubits, and consisting of $k$ layers, where each layer comprises parameterized $e^{i\theta X\otimes X}$ gates between neighboring qubits followed by a sequence of single qubit parameterized $Z$ and $X$ rotations:

\begin{figure}[h!]
\centering
\begin{quantikz}
\lstick{$\ket{0}$}&\gate[2]{e^{i\theta XX}} &\ghost{le}\qw &\gate{Rz(\theta)} &\gate{Rx(\theta)} &\gate[2]{e^{i\theta XX}} &\ghost{le}\qw &\gate{Rz(\theta)} &\gate{Rx(\theta)} &\qw \\
\lstick{$\ket{0}$}& &\gate[2]{e^{i\theta XX}} &\gate{Rz(\theta)} &\gate{Rx(\theta)} & &\gate[2]{e^{i\theta XX}} &\gate{Rz(\theta)} &\gate{Rx(\theta)} &\qw \\
\lstick{$\ket{0}$}&\qw & &\gate{Rz(\theta)} &\gate{Rx(\theta)} &\qw & &\gate{Rz(\theta)} &\gate{Rx(\theta)} &\qw 
\end{quantikz}\caption{A parameterized, layered quantum circuit. Each gate is dependent on a separate parameter, here all schematically represented as $\theta$.}
\end{figure}
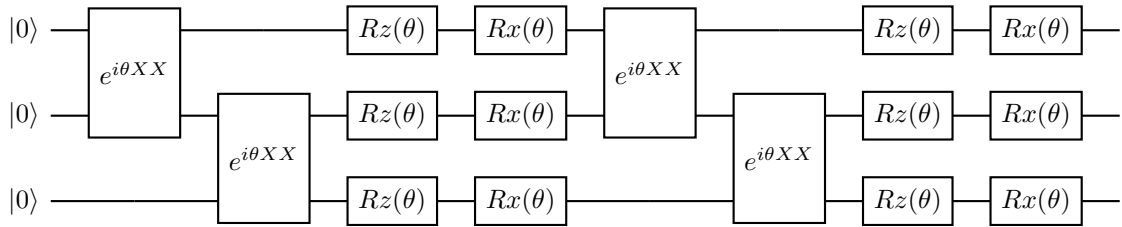
We now show how to implement such circuits in \tc, and how to use one of the machine learning backends to compute cost functions and gradients easily and efficiently.  
The circuit for general $n,k$ and set of parameters can be defined as follows:
\begin{lstlisting}[language=Python]
def qcircuit(n, k, params):
    c = tc.Circuit(n)
    for j in range(k):
        for i in range(n - 1):
            c.exp1(
                i, i + 1, theta=params[j * (3 * n - 1) + i], unitary=tc.gates._xx_matrix
            )
        for i in range(n):
            c.rz(i, theta=params[j * (3 * n - 1) + n - 1 + i])
            c.rx(i, theta=params[j * (3 * n - 1) + 2 * n - 1 + i])
    return c
\end{lstlisting}
As an example, we take $n=3, k=2$, set TensorFlow as our backend, and define an energy cost function to be minimized
$$E = \langle X_0 X_1\rangle_\theta + \langle X_1 X_2\rangle_\theta.$$ 

\begin{lstlisting}[language=Python]
n = 3
k = 2

K = tc.set_backend("tensorflow")


def energy(params):
    c = qcircuit(n, k, params)
    e = c.expectation_ps(x=[0, 1]) + c.expectation_ps(x=[1, 2])
    return K.real(e)
\end{lstlisting}

\bp{{\sf K.grad} and {\sf K.value\_and\_grad}.}
Using the ML backend support for automatic differentiation, we can now quickly compute both the energy and the gradient of the energy (with respect to the parameters): 
\begin{lstlisting}[language=Python]
energy_val_grad = K.value_and_grad(energy)
\end{lstlisting}
This creates a function which, given a set of parameters as input, returns both the energy and the gradient of the energy. If only the gradient is desired, then this can be computed by {\sf K.grad(energy)}. While we could run the above code directly on a set of parameters, if multiple evaluations of the energy will be performed, significant time savings can be had by using a just-in-time compiled version of the function:
\begin{lstlisting}[language=Python]
energy_val_grad_jit = K.jit(energy_val_grad)
\end{lstlisting}
With {\sf K.jit}, the initial evaluation of the energy and gradient may take longer, but subsequent evaluations will be noticeably faster than non-jitted code. We recommend always using {\sf jit} as long as the function is `tensor-in, tensor-out', and we have worked hard to make all aspects of the circuit simulator compatible with JIT.

\subsection{Optimization via ML backends}

With the energy function and gradients available, optimization of the parameters is straightforward.  Below is an example of how to do this via stochastic gradient descent:
\begin{lstlisting} [language=Python]
learning_rate = 2e-2
opt = K.optimizer(tf.keras.optimizers.SGD(learning_rate))


def grad_descent(params, i):
    val, grad = energy_val_grad_jit(params)
    params = opt.update(grad, params)
    if i % 10 == 0:
        print(f"i={i}, energy={val}")
    return params


params = K.implicit_randn(k * (3 * n - 1))
for i in range(200):
    params = grad_descent(params, i)
\end{lstlisting}
While this example was done with the TensorFlow backend, switching to JAX can be done easily. All that is required is to redefine the optimizer  {\sf opt} using the JAX optimization library optax \cite{optax}:
\begin{lstlisting}[language=Python]
import optax
opt = tc.backend.optimizer(optax.sgd(learning_rate)
\end{lstlisting}
Then, choose JAX as the backend via
\begin{lstlisting}[language=Python]
K = tc.set_backend("jax")
\end{lstlisting}
and perform the gradient descent exactly as above. Note that if no backend is set explicitly, \tc defaults to using NumPy as the backend, which does not allow for automatic differentiation. 

\subsection{Optimization via SciPy}
An alternative to using the machine learning backends for the optimization is to use SciPy.  This can be done via the {\sf scipy\_interface} API call, and allows for gradient based (e.g. BFGS) and non-gradient based (e.g. COBYLA) optimizers to be used, which are not available via the ML backends.

\begin{lstlisting}[language=Python]
import scipy.optimize as optimize

f_scipy = tc.interfaces.scipy_interface(energy, shape=[k * (3 * n - 1)], jit=True)
params = K.implicit_randn(k * (3 * n - 1))
r = optimize.minimize(f_scipy, params, method="L-BFGS-B", jac=True)
\end{lstlisting}
The first line above specifies the shape of the parameters to be supplied to the function to be minimized, which here is the energy function.  The {\sf jit=True} argument automatically takes care of jitting the energy function.  Gradient-free optimization can similarly be performed efficiently by supplying the {\sf gradient=False} argument to {\sf scipy\_interface}:
\begin{lstlisting}[language=Python]
f_scipy = tc.interfaces.scipy_interface(
    energy, shape=[k * (3 * n - 1)], jit=True, gradient=False
)
params = K.implicit_randn(k * (3 * n - 1))
r = optimize.minimize(f_scipy, params, method="COBYLA")
\end{lstlisting}

\section{Density matrices and mixed state evolution}

\begin{mdframed}
\textbf{Jupyter notebook: }
\fancylink{\rooturl docs/source/whitepaper/5-density-matrix.ipynb}{5-density-matrix.ipynb}
\end{mdframed}

\tc provides two methods for simulating noisy, mixed state quantum evolution.  Full density matrix simulation of $n$ qubits is provided by using {\sf tc.DMCircuit(n)}, and then adding quantum operations -- both unitary gates as well as general quantum operations specified by Kraus operators -- to the circuit.  Relative to pure state simulation of $n$ qubits via {\sf tc.Circuit}, full density matrix simulation is twice as memory intensive, and thus the maximum system size simulatable will be half of what can be simulated in the pure state case.  A less memory intensive option is to use the standard {\sf tc.Circuit(n)} object and stochastically simulate open system evolution via Monte Carlo methods.   

\subsection{Density matrix simulation with {\sf tc.DMCircuit}}

We illustrate this method below, by considering a simple circuit on a single qubit, which takes as input the mixed state corresponding to a probabilistic mixture of the $\ket{0}$ state and the maximally mixed state
\eq{
\rho(\alpha) = \alpha\outprod{0}{0} + (1-\alpha)I/2.
}
This state is then passed through a circuit which applies an $X$ gate, followed by a quantum operation corresponding to an amplitude damping channel $\mathcal{E}_\gamma$ with parameter $\gamma$. This has Kraus operators
\eq{
K_0 = \begin{pmatrix}
1 & 0 \\ 0 & \sqrt{1-\gamma}
\end{pmatrix}, \quad K_1 = \begin{pmatrix}
0 & \sqrt{\gamma} \\ 0 & 0
\end{pmatrix}
}
This circuit thus induces the evolution
\eq{
\rho(\alpha) \xrightarrow[]{X} X\rho(\alpha)X\xrightarrow[]{\mathcal{E}_\gamma}\sum_{i=0}^1 K_i X\rho(\alpha)X K_i^\dagger
}
To simulate this in \tc, we first create a {\sf tc.DMCircuit} (density matrix circuit) object and set the input state using the {\sf dminputs} optional argument (note that if a pure state input is provided to {\sf tc.DMCircuit}, this should be done via the {\sf inputs} optional argument rather than {\sf dminputs}). $\rho(\alpha)$ has the matrix form
\eq{
\rho(\alpha) = \begin{pmatrix}
\frac{1+\alpha}{2} & \\ & \frac{1-\alpha}{2}
\end{pmatrix},
}
and thus (taking $\alpha=0.6$) we initialize the density matrix circuit as follows 
\begin{lstlisting}[language=Python]
def rho(alpha):
    return np.array([[(1 + alpha) / 2, 0], [0, (1 - alpha) / 2]])


input_state = rho(0.6)
dmc = tc.DMCircuit(1, dminputs=input_state)
\end{lstlisting}
Adding the $X$ gate (and other unitary gates) is done in the same way as for pure state circuits:
\begin{lstlisting}[language=Python]
dmc.x(0)                  
\end{lstlisting}
To implement a general quantum operation such as the amplitude damping channel, we use {\sf general\_kraus}, supplied with the corresponding list of Kraus operators.
\begin{lstlisting}[language=Python]
def amp_damp_kraus(gamma):
    K0 = np.array([[1, 0], [0, np.sqrt(1 - gamma)]])
    K1 = np.array([[0, np.sqrt(gamma)], [0, 0]])
    return K0, K1


K0, K1 = amp_damp_kraus(0.3)
dmc.general_kraus([K0, K1], 0)  # apply channel with Kraus operators [K0,K1] to qubit 0 
\end{lstlisting}
and the full density matrix output can be returned via
\begin{lstlisting}[language=Python]
dmc.state()                 
\end{lstlisting}
In this example we input the Kraus operators for the amplitude damping channel manually, in order to illustrate the general approach to implementing quantum channels with the {\sf tc.DMCircuit} method. In fact, \tc includes built-in methods for returning the Kraus operators for a number of common channels, including the amplitude damping, depolarizing, phase damping and reset channels. For example, the Kraus operators for the phase damping channel with parameter $\gamma$
\eq{
K_0 = \begin{pmatrix}
1 & 0 \\ 0 & \sqrt{1-\gamma}
\end{pmatrix}, \quad K_1=\begin{pmatrix}
0 & 0 \\ 0 & \sqrt{\gamma}
\end{pmatrix}
}
can be returned by calling
\begin{lstlisting}[language=Python]
gamma = 0.3
K0, K1 = tc.channels.phasedampingchannel(gamma) 
\end{lstlisting}
and the phase damping channel added to a circuit via
\begin{lstlisting}[language=Python]
dmc.general_kraus([K0, K1], 0)
\end{lstlisting}
The above operation can be further simplified using one API call:
\begin{lstlisting}[language=python]
dmc.phasedamping(0, gamma=0.3)
\end{lstlisting}

\subsection{Monte Carlo simulation with {\sf tc.Circuit}}\label{sec:MCsim}
Monte Carlo methods can be used to sample noisy quantum evolution using {\sf tc.Circuit} instead of {\sf tc.DMCircuit}, where the mixed state is effectively simulated by an ensemble of pure states.
As for density matrix simulation, quantum channels $\mathcal{E}$ can be added to a {\sf circuit} object by providing a list of their associated Kraus operators $\{K_i\}$.  The API is the same as for the full density matrix simulation:
\begin{lstlisting}[language=Python]
input_state = np.array([1, 1] / np.sqrt(2))
c = tc.Circuit(1, inputs=input_state)
c.general_kraus(tc.channels.phasedampingchannel(0.5), 0)
c.state()
\end{lstlisting}
In this framework though, the output of a channel acting on $\ket{\psi}$ , i.e.
\eq{
\mathcal{E}\lp \ket{\psi}\bra{\psi}\rp &= \sum_i K_i \outprod{\psi}{\psi} K_i^\dag 
}
is viewed as an ensemble of states $\frac{K_i\ket{\psi}}{\sqrt{\bra{\psi}K_i^\dag K_i \ket{\psi}}}$ that each occurs with probability $p_i = \bra{\psi}K_i^\dag K_i \ket{\psi}$.  Thus, the code above stochastically produces the output of a single qubit initialized in state $\ket{\psi}=\frac{\ket{0}+\ket{1}}{\sqrt{2}}$ being passed through a phase damping channel with parameter $\gamma=0.5$.  

The Monte Carlo simulation of channels where the Kraus operators are all unitary matrices (up to a constant factor) can be more efficiently handled by using {\sf unitary\_kraus} instead of {\sf general\_kraus}. For instance,  the depolarizing channel with Kraus operators parameterized by $p_x, p_y,p_z$:
\eq{
K_0 = (1-p_x-p_y-p_z) \begin{pmatrix}
1 & 0 \\ 0 & 1
\end{pmatrix}, \quad K_1 = p_x \begin{pmatrix}
0 & 1 \\ 1 & 0
\end{pmatrix}, \quad K_2 = p_y\begin{pmatrix}
0 & -i \\ i & 0
\end{pmatrix}, \quad K_3 = p_z \begin{pmatrix}
1 &0 \\0 & -1
\end{pmatrix}
}
can be implemented by
\begin{lstlisting}[language=Python]
px, py, pz = 0.1, 0.2, 0.3
c.unitary_kraus(tc.channels.depolarizingchannel(px, py, pz), 0)
\end{lstlisting}
where, in the second line, {\sf tc.channel.depolarizingchannel(px, py, pz)} returns the required Kraus operators.

\subsubsection{Externalizing the randomness} \label{sec:vmap_MC}
The {\sf general\_kraus} and {\sf unitary\_kraus} examples above both handle randomness generation from inside the respective methods. That is, when the list $[K_0, K_1, \ldots, K_{m-1}]$ of Kraus operators is supplied to {\sf general\_kraus} or {\sf unitary\_kraus}, the method  partitions the interval $[0,1]$ into $m$ contiguous intervals $[0,1] = I_0 \cup I_1 \cup \ldots I_{m-1}$ where the length of $I_i$ is proportional to the relative probability of obtaining the outcome $i$. Then a uniformly random variable $x$ in $[0,1]$ is drawn from within the method, and outcome $i$ is selected based on which interval $x$ lies in. In \tc, we have full backend agnostic infrastructure for random number generation and management. However, the interplay between jit, random numbers and backend switching is often subtle if we rely on the random number generation inside these methods. See \fancylink{\docurl advance.html\#randoms-jit-backend-agnostic-and-their-interplay}{advance.html\#randoms-jit-backend-agnostic-and-their-interplay} for details.

In some situations, it may be preferable to first generate the random variable from outside the method, and then pass the value generated into {\sf general\_kraus} or {\sf unitary\_kraus}.  This can be done via the optional {\sf status} argument:
\begin{lstlisting}[language=Python]
px, py, pz = 0.1, 0.2, 0.3
x = K.implicit_randn()
c.unitary_kraus(tc.channels.depolarizingchannel(px, py, pz), 0, status=x)
\end{lstlisting}
This is useful, for instance, when one wishes to use {\sf vmap} to batch compute multiple runs of a Monte Carlo simulation. This is illustrated in the example below, where {\sf vmap} is used to compute 10 runs of the simulation in parallel:
\begin{lstlisting}[language=Python]
def f(x):
    c = tc.Circuit(1)
    c.h(0)
    c.unitary_kraus(tc.channels.depolarizingchannel(0.1, 0.2, 0.3), 0, status=x)
    return c.state()


f_vmap = K.vmap(f, vectorized_argnums=0)
X = K.implicit_randn(10)
f_vmap(X)
\end{lstlisting}
Conceptually, the line
\begin{lstlisting}[language=Python]
f_vmap = K.vmap(f, vectorized_argnums=0)
\end{lstlisting}
creates a function which acts as
\eq{
f_{vmap}\begin{pmatrix} x_0 \\ x_1 \\ \vdots 
\end{pmatrix}
 &= \begin{pmatrix}
 f(x_0) \\ f(x_1) \\  \vdots
 \end{pmatrix}
}
and the argument {\sf vectorized\_argnums=0} indicates that is the zeroth argument (in this case the only argument) of $f$ that we wish to batch compute in parallel.

\section{Advanced features}

\markedup{This section consists of advanced features that the general reader may skip on first reading.  The reader interested in the benchmark studies of Section~\ref{sec:examples_benchmarks} may wish to refer to this section though, as those examples use a number of concepts from here.}

\subsection{Conditional measurements and post-selection}\label{sec:cond-measure}

\begin{mdframed}
\textbf{Jupyter notebook: }
\fancylink{\rooturl docs/source/whitepaper/6-1-conditional-measurements-post-selection.ipynb}{6-1-conditional-measurements-post-selection.ipynb}
\end{mdframed}

\tc allows for two kinds of operations to be performed that are related to  measurement outcomes.  These are (i) conditional measurements, the outcomes of which can be used to control downstream conditional quantum gates, and (ii) post-selection, which allows the user to select the post-measurement state corresponding to a particular measurement outcome.

\subsubsection{Conditional measurements}

The {\sf cond\_measure} command is used to simulate the process of performing a Z measurement on a qubit,  generating a measurement outcome with probability given by the Born rule, and collapsing the wavefunction in accordance with the measured outcome.  The classical measurement outcome obtained can then act as a control for a subsequent quantum operation via the {\sf conditional\_gate} API and can be used, for instance, to implement the canonical teleportation circuit.  

\begin{lstlisting}[language=Python]
# quantum teleportation of state |psi> = a|0> + sqrt(1-a^2)|1>
a = 0.3
input_state = np.kron(np.array([a, np.sqrt(1 - a ** 2)]), np.array([1, 0, 0, 0]))

c = tc.Circuit(3, inputs=input_state)
c.h(2)
c.cnot(2, 1)
c.cnot(0, 1)
c.h(0)

# mid-circuit measurements
z = c.cond_measure(0)
x = c.cond_measure(1)

# if x = 0 apply I, if x = 1 apply X (to qubit 2)
c.conditional_gate(x, [tc.gates.i(), tc.gates.x()], 2)

# if z = 0 apply I, if z = 1 apply Z (to qubit 2)
c.conditional_gate(z, [tc.gates.i(), tc.gates.z()], 2)
\end{lstlisting}

\begin{figure}[h!]
\centering
\begin{quantikz}
\lstick{$\ket{\psi}$}&\qw &\qw &\ctrl{1} &\gate{H} &\meter{}  &\cwbend{2} \\
\lstick{$\ket{0}$}&\qw &\targ{} &\targ{} &\qw &\meter{} \vcw{1}  \\
\lstick{$\ket{0}$}&\gate{H} &\ctrl{-1}  &\qw &\qw &\gate{X} &\gate{Z} 
\end{quantikz}\caption{Teleportation circuit implemented with {\sf c.cond\_measure} and {\sf c.conditional\_gate}.}\label{fig:teleportation-circuit}
\end{figure}
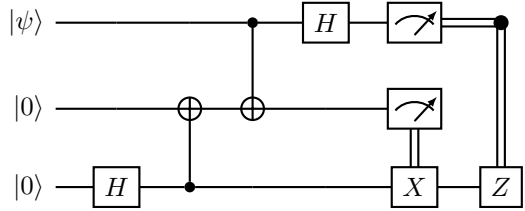

\subsubsection{Post-selection}
Post-selection is enabled in \tc via the {\sf post\_select} method.  This allows the user to select the post-$Z$-measurement state of a qubit via the {\sf keep} argument. Unlike {\sf cond\_measure}, the state returned by {\sf post\_select} is not normalized. As an example, consider
\begin{lstlisting}[language=Python]
c = tc.Circuit(2, inputs=np.array([1, 0, 0, 1] / np.sqrt(2)))
c.post_select(0, keep=1)  # measure qubit 0, post-select on outcome 1
c.state()                 
\end{lstlisting}
which initializes a $2$-qubit maximally entangled state $\ket{\psi} = \frac{\ket{00}+\ket{11}}{\sqrt{2}}$. The first qubit ($q_0$) is then measured in the $Z$-basis, and the unnormalized state  $\ket{11}/\sqrt{2}$ corresponding to the measurement outcome $1$ being post-selected. 

This post-selection scheme with unnormalized states is fast and can, for instance, be used to explore various quantum algorithms and nontrivial quantum physics such as measurement-induced entanglement phase transitions \cite{PhysRevB.98.205136,PhysRevB.99.224307,PhysRevX.9.031009, PhysRevB.100.134306, arXiv:2212.03901}.

\subsection{Pauli string expectation}\label{sub:pss}

\begin{mdframed}
\textbf{Jupyter notebook: }
\fancylink{\rooturl docs/source/whitepaper/6-2-pauli-string-expectation.ipynb}{6-2-pauli-string-expectation.ipynb}
\end{mdframed}

Minimizing the expectation values of sums of Pauli strings is a common task in quantum algorithms. For instance, in the VQE ground state preparation of an $n$-site transverse-field Ising model (TFIM) with Hamiltonian 
\eq{
H &= \sum_{i=0}^{n-2} J_i X_i X_{i+1} - \sum_{i=0}^{n-1} h_i Z_i,
}
where the $J_i,h_i$ are model parameters, one wishes to minimize
\eql{
E(\theta)=\langle H \rangle_\theta:= \sum_{i=0}^{n-2} J_i\langle X_i X_{i+1}\rangle_\theta- \sum_{i=0}^{n-1} h_i\langle Z_i\rangle_\theta \label{eq:tfim}
}
with respect to the circuit parameters $\theta$. \tc provides a number of ways to compute expressions of this form, useful in different scenarios.  At a high level these are
\begin{itemize}
    \item Looping over terms, each computed by {\sf c.expectation\_ps}.
    \item Supplying a dense matrix, sparse matrix or MPO representation of the Hamiltonian to the {\sf operator\_expectation} function.
    \item Using {\sf vmap} to compute each of the terms in a vectorized parallel manner, where each term is input as a \textit{structure} vector.
\end{itemize}
Underlying these methods are a variety of ways of representing strings of Pauli operators and converting between these representations. Before going through the above methods in more detail, let us introduce the Pauli structure vector representation utilized in \tc.

\subsubsection{Pauli structures and weights}\label{sec:structures}
A string of Pauli operators acting on $n$ qubits can be represented as a length-$n$ vector $v\in\{0,1,2,3\}^n$, where the value of $v_i = j$ corresponds to $\sigma_i^j$, i.e. Pauli operator $\sigma^j$ acting on qubit $i$ (with $\sigma^0 = I, \sigma^1 = X, \sigma^2 = Y, \sigma^3 = Z$). For example, in this notation, if $n=3$ the term $X_1 X_2$ corresponds to $v = [0,1,1]$. We refer to such a vector representation of a Pauli string as a \textit{structure}, and a list of structures, one for each Pauli string term in the Hamiltonian, is used as the input to compute sums of expectation values in a number of ways.
\begin{lstlisting}[language=Python]
# Pauli structures for Transverse Field Ising Model
structures = []
for i in range(n - 1):
    s = [0 for _ in range(n)]
    s[i] = 1
    s[i + 1] = 1
    structures.append(s)
for i in range(n):
    s = [0 for _ in range(n)]
    s[i] = 3
    structures.append(s)
\end{lstlisting}
If each structure has an associated weight, e.g. the term $X_i X_{i+1}$ has weight $J_i$ in Hamiltonian~\eqref{eq:tfim}, then we define a corresponding tensor of weights
\begin{lstlisting}[language=Python]
# Weights, taking J_i = 1.0, all h_i = -1.0
J_vec = [1.0 for _ in range(n - 1)]
h_vec = [-1.0 for _ in range(n)]
weights = tc.array_to_tensor(np.array(J_vec + h_vec))
\end{lstlisting}

\subsubsection{Explicit loop with {\sf c.expectation\_ps}} \label{sub:explicitloop}
As introduced in Section~\ref{sec:basics}, given a \tc quantum circuit $c$, a single Pauli string expectation can be computed by supplying a list of indices to {\sf c.expectation\_ps}. The sum can then be computed by using a straightforward loop:
\begin{lstlisting}[language=Python]
def tfim_energy(c,J_vec, h_vec)
    e = 0.0
    n = c._nqubits
    for i in range(n):
        e+= h_vec[i] * c.expectation_ps(z=[i])
    for i in range(n-1):
        e+= J_vec[i] * c.expectation_ps(x=[i,i+1])
    return K.real(e)
\end{lstlisting}

\subsubsection{Expectations of Hamiltonians via {\sf operator\_expectation}}  \label{sub:mporep}
Given a \tc quantum circuit {\sf c} and an operator {\sf op} representing the Hamiltonian, the expectation value of the energy can also be computed via the {\sf operator\_expectation} API from the \tc templates library
\begin{lstlisting}[language=Python]
e =  tc.templates.measurements.operator_expectation(c, op)
\end{lstlisting}
The operator {\sf op} itself can be expressed in one of three forms: (i) dense matrix, (ii) sparse matrix, and (iii) Matrix Product Operator (MPO). 

\bp{Dense Matrix Input.} As a simple example, take the Hamiltonian
\eq{
X_0X_1  - Z_0 - Z_1.
}
This has dense matrix representation
\eq{
\begin{pmatrix}
-2 & & & 1 \\
& & 1& \\
& 1 & & \\
1 & & &2
\end{pmatrix}
}
and the expectation value of the Hamiltonian
(with respect to circuit {\sf c}) can be computed by
\begin{lstlisting}[language=Python]
op = tc.array_to_tensor([[-2, 0, 0, 1], [0, 0, 1, 0], [0, 1, 0, 0], [1, 0, 0, 0]])
e = tc.templates.measurements.operator_expectation(c, op)
\end{lstlisting}
The matrix elements above were input by hand.  \tc also provides a way of generating the matrix elements from the associated Pauli structures and weights, e.g.
\begin{lstlisting}[language=Python]
structure = [[1, 1], [0, 3], [3, 0]]
weights = [1.0, -1.0, -1.0]
H_dense = tc.quantum.PauliStringSum2Dense(structure, weights)
\end{lstlisting}

\bp{Sparse Matrix Input.} Significant computational advantage in terms of space and time can be obtained if the Hamiltonian is sparse, in which case a sparse representation of the operator is preferable. This can be implemented in a backend agnostic way by converting from a list of Pauli structures in a two-stage process. First we convert to a sparse numpy matrix in COO (COOrdinate) format. For example, with the structures and weights defined in Section~\ref{sec:structures}, we call
\begin{lstlisting}[language=Python]
H_sparse_numpy = tc.quantum.PauliStringSum2COO_numpy(structures, weights)
\end{lstlisting}
Then we can convert to a sparse tensor compatible with the selected backend {\sf K}
\begin{lstlisting}[language=Python]
H_sparse = K.coo_sparse_matrix(
    np.transpose(np.stack([H_sparse_numpy.row, H_sparse_numpy.col])),
    H_sparse_numpy.data,
    shape=(2 ** n, 2 ** n),
)
\end{lstlisting}
and then call {\sf operator\_expectation} on this sparse tensor
\begin{lstlisting}[language=Python]
e = tc.templates.measurements.operator_expectation(c, H_sparse)
\end{lstlisting}

\bp{MPO Input.}  The TFIM Hamiltonian, as a short-ranged spin Hamiltonian, admits an efficient Matrix Product Operator representation. Again this is a two-stage process using \tc. We first convert the Hamiltonian into an MPO representation via the TensorNetwork~\cite{tensornetwork} or Quimb package~\cite{quimb}:
\begin{lstlisting}[language=Python]
# generate the corresponding MPO by converting the MPO in tensornetwork package

Jx = np.array([1.0 for _ in range(n - 1)])  # strength of xx interaction (OBC)
Bz = np.array([1.0 for _ in range(n)])  # strength of transverse field
# Note the convention for the sign of Bz
hamiltonian_mpo = tn.matrixproductstates.mpo.FiniteTFI(Jx, Bz, dtype=np.complex64)
\end{lstlisting}
and then convert the MPO into a QuOperator object compatible with \tc.  

\begin{lstlisting}[language=Python]
hamiltonian_mpo = tc.quantum.tn2qop(hamiltonian_mpo)  # QuOperator in TensorCircuit
\end{lstlisting}
The expectation value of the energy can then be computed via {\sf operator\_expectation}
\begin{lstlisting}[language=Python]
e = tc.templates.measurements.operator_expectation(c, hamiltonian_mpo)
\end{lstlisting}

\subsubsection{{\sf vmap} over Pauli structures}\label{sec:vmap_pauli}

Given a state $\ket{s}$, the expectation value $\bra{s} P \ket{s}$ of a Pauli string $P$ with a given structure can be computed as a function of tensor inputs as
\begin{lstlisting}[language=Python]
# assume the following are defined
# state: 2**n vector of coefficients corresponding to input state
# structure: Pauli structure (i.e. list of integers {0,1,2,3}**n)

structure = tc.array_to_tensor(structure)
state = tc.array_to_tensor(state)


def e(state, structure):
    c = tc.Circuit(n, inputs=state)
    return tc.templates.measurements.parameterized_measurements(
        c, structure, onehot=True
    )
\end{lstlisting}
where the {\sf parameterized\_measurements} function is used to compute the expectation of the output of a circuit. Then, if the Hamiltonian is represented by a list of Pauli structures $[v_1, \ldots, v_k]$, {\sf vmap} can be used to compute the expectation with respect to the circuit {\sf c} of each term in parallel:
\begin{lstlisting}[language=Python]
e_vmap = K.vmap(e, vectorized_argnums=1)
\end{lstlisting}
Similar to the example in Section~\ref{sec:vmap_MC}, vmapping creates a function which acts as
\eq{
e_{vmap}\lp s, \begin{pmatrix} \leftarrow v_1 \rightarrow\\ \vdots \\ \leftarrow v_k \rightarrow\end{pmatrix}\rp &=
\begin{pmatrix} e(s,v_1)\\ \vdots \\
e(s,v_k)\end{pmatrix}
}
i.e., outputs a vector of expectation values corresponding to terms in the Hamiltonian, and 
\begin{lstlisting}[language=Python]
 vectorized_argnums=1
\end{lstlisting}
indicates that it is the first argument {\sf v} of the {\sf e(s,v)} function which should be computed in parallel, while the zeroth argument (i.e. $s$) is fixed.  With structures and weights as defined in Section~\ref{sec:structures}, the expectation value of the Hamiltonian with respect to the circuit $c$ can then be computed as
\begin{lstlisting}[language=Python]
s = c.state()
e_terms = e_vmap(s, structures)
hamiltonian_expectation = K.sum(e_terms * K.real(weights))
\end{lstlisting}

We benchmark these different approaches to Pauli string sum evaluation for Hamiltonians corresponding to an H$_2$O molecule and the TFIM spin model, with results in Tables~\ref{tab:h2ops} and~\ref{tab:tfimps}, respectively. See \fancylink{\rooturl examples/vqeh2o_benchmark.py}{examples/vqeh2o\_benchmark.py} and \fancylink{\rooturl examples/vqetfim_benchmark.py}{examples/vqetfim\_benchmark.py} for more details. In the H$_2$O case, we observe an acceleration of {\emph {85,000 times}} for VQE evaluation using the sparse matrix representation compared to a na\"ive loop \markedup{(also implemented in \tc)} as used in most other quantum software. 

\bp{General remark on benchmark times:} \markedup{Throughout this work, when averages times per circuit or gradient evaluation are reported (either using \tc or other software), the initial compilation time associated with JIT is not reported.  In practice, we find that this initial compilation time is typically an order of ten to several tens larger than the time required for subsequent evaluations, and its effect on the total algorithmic running time depends on how many iterations it is amortized over.  If a circuit is only evaluated a few times, the time required for JIT compilation may outweigh the benefits it brings.  However, in typical variational quantum algorithms or in noise simulation based on Monte Carlo trajectories, circuits are evaluated a large number of times, and the amortized cost of JIT compilation is negligible.} 

%Throughout this work, when benchmark times are reported per iteration or per gradient evaluation actual running time where the possible compiling time for packages with JIT mechanism is amortized and omitted. In other words, comparing to other quantum simulator solutions, \tc has no unique advantage to simulate generic circuit once where the compiling time cannot be amortized. Each circuit is evaluated multiple times during variational algorithm optimization loops, or in noise simulation based on Monte Carlo trajectories. \tc greatly outperforms other quantum packages in these NISQ scenarios and more benchmark results are presented in the following sections.

\begin{table}\centering
\begin{tabular}{c|c|c}
  time (s)   & CPU & GPU  \\ 
     \hline 
explicit loop  &   65.7    &    119   \\
vmap over Pauli structures  &   0.68    &  0.0018   \\
dense matrix representation   &   0.26    &   0.0015      \\
sparse matrix representation  &   ~0.008  ~  &   ~ 0.0014    ~  \\

\end{tabular}
\caption{Performance benchmarks for Pauli string sum evaluation methods, with Hamiltonian corresponding to an H$_2$O molecule. 12 qubits are used for the binary encoding of STO-3G orbitals. The qubit Hamiltonian contains 1390 Pauli string terms in total. Data was obtained using the JAX backend. GPU simulations were performed on the Nvidia T4 GPU, while CPU simulations used Intel(R) Core(TM) i7-9750H CPU @ 2.60GHz. The MPO representation is not applicable in this case, since the required bond dimension is too large.}
\label{tab:h2ops}
\end{table}

\begin{table}\centering
\begin{tabular}{c|c|c}
  time (s)   & CPU & GPU  \\ 
     \hline 
explicit loop  &   1.73    &0.11   \\
vmap over Pauli structures  &  10.68   &  0.20  \\
sparse matrix representation   &   0.61    &   0.0086      \\
MPO matrix representation  &   ~0.0007 ~  &   ~ 0.0039    ~  \\

\end{tabular}
\caption{Performance benchmark for different Pauli string sum evaluation on a 20-qubit TFIM Hamiltonian with open boundary conditions. The qubit Hamiltonian contains 39 Pauli string terms. Data was obtained using the JAX backend. GPU  simulations used the Nvidia T4 GPU, while CPU  simulations used Intel(R) Xeon(R) Platinum 8255C CPU @ 2.50GHz. The dense matrix representation is not applicable for systems of 20 qubits due to excessive memory requirements.}
\label{tab:tfimps}
\end{table}

\subsection{{\sf vmap} and {\sf vectorized\_value\_and\_grad}}\label{sec:vmap}

\begin{mdframed}
\textbf{Jupyter notebook: }
\fancylink{\rooturl docs/source/whitepaper/6-3-vmap.ipynb}{6-3-vmap.ipynb}
\end{mdframed}

As we have seen in Sections~\ref{sec:vmap_MC} and~\ref{sec:vmap_pauli}, {\sf vmap} allows for batches of function evaluations to be performed simultaneously in parallel. 
If batch evaluation of gradients as well as function values is required, then this can be done via ${\sf vectorized\_value\_and\_grad}$. In the simplest case, consider a function 
$f(x,y)$ where $x\in\mb{R}^p,y\in\mb{R}^q$ are both vectors, and one wishes to evaluate both $f(x,y)$ and $\sum_x\nabla_y f(x,y) = \sum_x\lp \frac{\pd f(x,y_1)}{\pd y_1},\ldots, \frac{\pd f(x,y_q)}{\pd y_q}\rp^\top$ over a batch $x_1, x_2,\ldots, x_k$ of inputs $x$. This is achieved by creating a new, vectorized value-and-gradient function 
\eq{
f_{vvg}\lp \begin{pmatrix} \leftarrow x_1 \rightarrow\\ \vdots \\ \leftarrow x_k \rightarrow\end{pmatrix}, y\rp &=
\begin{pmatrix} \begin{pmatrix}f(x_1, y) \\ \vdots \\
f(x_k,y)\end{pmatrix},\sum_{i=1}^k \nabla_y f(x_i,y) \end{pmatrix}
}
which takes as zeroth argument the batched inputs expressed as a $k\times p$ tensor, and as first argument the variables we wish to differentiate with respect to. The outputs are a vector of function values evaluated at all points $(x_i,y)$, and the gradient averaged over all those points. A toy example is implemented as follows: 
 
\begin{lstlisting}[language=Python]
def f(x, y):
    return x[0] * x[1] * y[0] ** 2


f_vvg = K.vectorized_value_and_grad(f, argnums=1, vectorized_argnums=0)
X = tc.array_to_tensor([[1, 2], [2, 3], [0, -1]])
y = tc.array_to_tensor([2])
f_vvg(x_tensor, y)
\end{lstlisting}
{\sf argnums} indicates which argument we wish to take derivatives with respect to, and {\sf vectorized\_argnums} indicates which argument corresponds to the batched input.  It is even possible to set the values of {\sf argnums} and {\sf vectorized\_argnums} to be the same, i.e., we batch compute over different initial values of the parameters we wish to optimize over. This can be useful, for example, in batched VQE computations (see Section~\ref{sec:batched_vqe}).

\subsubsection{Batched input states}

Consider a quantum circuit $U(w)$ on $n$ qubits,  parameterized by weights $w = [w_1, \ldots, w_k]$, and with state-dependent loss function $f(\psi, w)$, e.g.
\eq{
\mathcal{L}=\sum_{\psi}f(\psi,w) = \sum_{\psi}\bra{\psi}U^\dag(w) Z_2 U(w)\ket{\psi}
}
Then, both the value $f(\psi,w)$ and the gradient with respect to the weights $\nabla_wf(\psi,w)$ 
can be evaluated for a batch of $k$ input states $\ket{\psi_1}, \ldots, \ket{\psi_k}$  simultaneously. 

\begin{figure}[htp]
    \centering
    \includegraphics[width=0.7\textwidth]{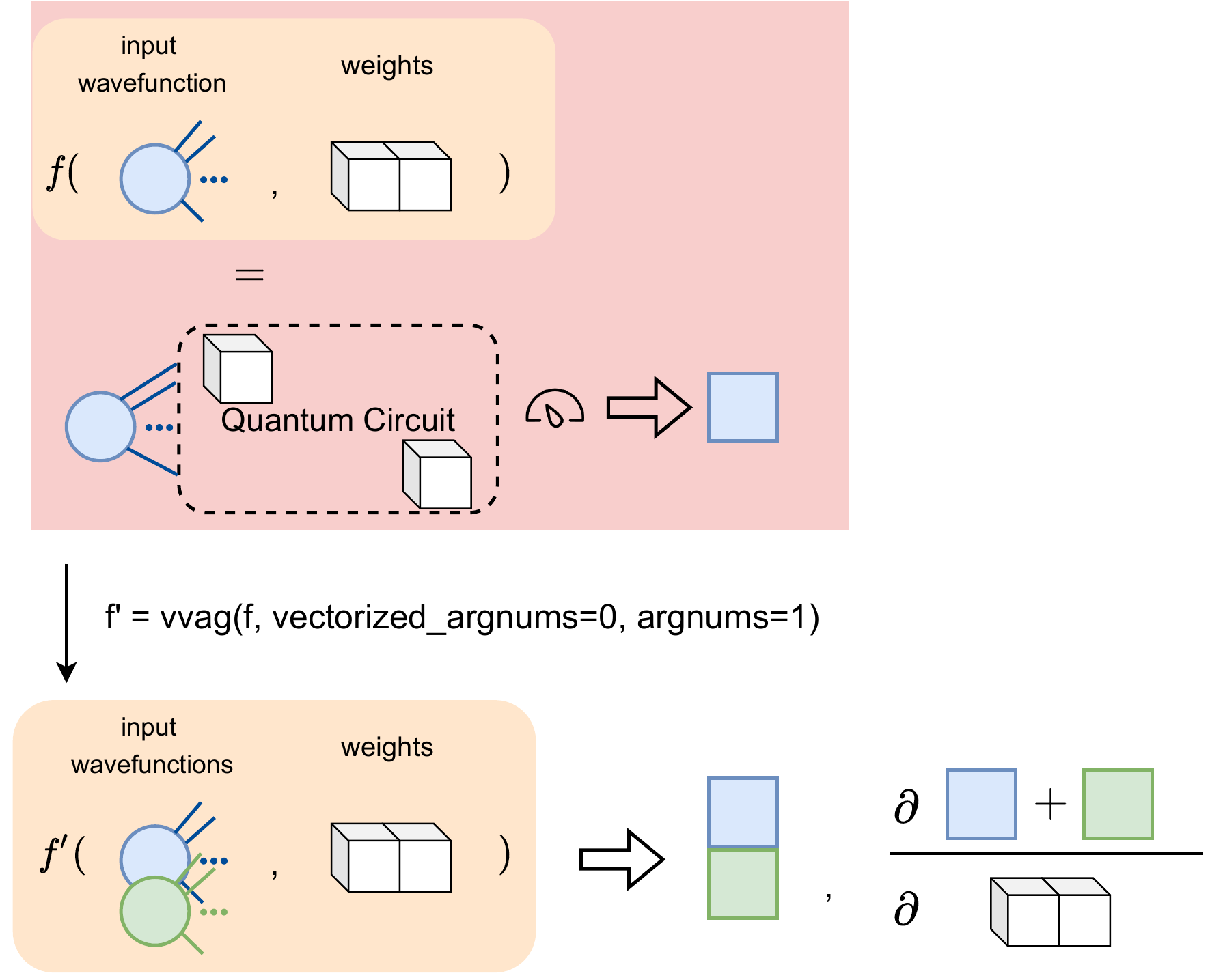}

\caption{Schematic of applying {\sf vectorized\_value\_and\_grad} on quantum simulation tasks with multiple input states. In the figure, the function f' at the bottom is transformed from the original function f defined at the top via {\sf f' = K.vectorized\_value\_and\_grad(f)}. The {\sf wavefunction} input is vectorized while differentiation is with respect to the {\sf weights} input. }
    \label{fig:qml}
\end{figure}

\begin{lstlisting}[language=Python]
f_vvg = K.vectorized_value_and_grad(f, argnums=1, vectorized_argnums=0)
f_vvg(psi_matrix, weights)
\end{lstlisting}
The vectorized value and grad workflow for batched input states can be visualized in Figure~\ref{fig:qml}.

\subsubsection{Batched circuits}\label{sub:vmapcircuit}

Consider a family of parameterized quantum circuits $U_1(w), \ldots, U_k(w)$ acting on the same number of qubits, where each circuit $U_i(w)$ can be expressed as a different parametrization of a single parent circuit, i.e.
\eq{
U_j(w) = U(w, x_j)
}
where $x_j$ is a vector of parameters. A loss function $f$ is defined on these circuits, e.g. $f(w,x_j) = \bra{0}U_j^\dag X_1U_j\ket{0}$ and its gradient with respect to the weights $w$ can also be batch evaluated across all circuits simultaneously, e.g. by defining
\eq{
f_{vvg}\lp w, \begin{pmatrix} \leftarrow x_1 \rightarrow\\ \vdots \\ \leftarrow x_k \rightarrow\end{pmatrix}\rp &=
\begin{pmatrix}\begin{pmatrix} f(w,x_1) \\ \vdots \\
f(w,x_k)\end{pmatrix}, \sum_{i=1}^k\nabla_wf(w,x_i)\end{pmatrix}
}
In this case, the batched inputs correspond to the first argument, and the variables to differentiate with respect to correspond to the zeroth argument:
\begin{lstlisting}[language=Python]
f_vvg = K.vectorized_value_and_grad(f, argnums=0, vectorized_argnums=1)
f_vvg(weights,x_matrix)
\end{lstlisting}

\subsubsection{Batched cost function evaluation}

As discussed in Section~\ref{sec:vmap_pauli}, {\sf vmap} can be used to accelerate the computation of expectations of sums of Pauli strings by making use of the Pauli structure vectors representation.  The same can be done for gradients of such expectation values using {\sf vectorized\_value\_and\_grad}. Consider a circuit on $n=3$ qubits, with $k=4$ Pauli terms and cost function
\eq{
f(w) = \langle Z_0 X_2\rangle + \langle X_1 Y_2\rangle + \langle X_0 X_1 Z_2\rangle +\langle Z_1\rangle
}
where $\langle Z_0 X_2 \rangle : = \bra{0}U^\dag(w) Z_0 X_2 U(w)\ket{0}$ for some parameterized circuit $U(w)$, and similarly for the other terms.  Suppose we have a parameterized circuit defined:
\begin{lstlisting}[language=Python]
def param_circuit(params):
    c = tc.Circuit(n)
    # circuit details omitted
    return c
\end{lstlisting}
then, we can express the expectation of a Pauli term with structure $v$ as:
\begin{lstlisting}[language=Python]
def f(w, v):
    c = param_circuit(w)
    return tc.templates.measurements.parameterized_measurements(c, v, onehot=True)
\end{lstlisting}
To batch compute the value and gradient of each term in the cost function, we create a tensor of the Pauli structures:
\begin{lstlisting}[language=Python]
structures = tc.array_to_tensor([[3,0,1], #<Z0 X2>
                                 [0,1,2], #<X1 Y2>
                                 [1,1,3], #<X0 X1 Z2>
                                 [0,3,0]]) #<Z1>

\end{lstlisting}
which we then pass in to the {\sf vectorized\_value\_and\_grad} version of $f$
\begin{lstlisting}[language=Python]
f_vvg = K.vectorized_value_and_grad(f, argnums=0, vectorized_argnums=1)
f_vvg(params, structures)
\end{lstlisting}
Schematically, this returns
\eq{
f_{vvg}\lp w, \begin{pmatrix} \leftarrow v_1 \rightarrow\\ \vdots \\ \leftarrow v_k \rightarrow\end{pmatrix}\rp &=
\begin{pmatrix}\begin{pmatrix} f(w,v_1), \\ \vdots \\
f(w,v_k)\end{pmatrix}, \sum_{i=1}^k\nabla_wf(w,v_i)\end{pmatrix}
}
with each vector $v_i$ corresponding to a different Pauli term. 

\subsubsection{Batched Machine Learning}
In machine learning problems, one often wishes to perform batch computation over $(\text{data}, \text{label})$ pairs.  This can be done by supplying a tuple of indices to the {\sf vectorized\_argnums} argument of {\sf vectorized\_value\_and\_grad}.  

Consider the following toy problem where data vectors $x\in[0,1]^2$ have labels $y\in\{0,1\}$ and one wishes to train the weights of a parameterized quantum circuit based on a training set of $(x,y)$ pairs.  The $x$ vectors are encoded in the angles of an initial set of parameterized gates, with the remaining weights $w$ to be trained to minimize the cost function below:
\begin{lstlisting}[language=Python]
def f(x, y, w):
    c = tc.Circuit(2)

    # encode x in qubit rotations
    c.rx(0, theta=x[0])
    c.rx(1, theta=x[1])

    # # parameterized circuit to determine label
    c.rx(0, theta=w[0])
    c.cnot(0, 1)
    c.rx(1, theta=w[1])

    yp = c.expectation_ps(z=[1])
    return K.real(e - y[0]) ** 2, yp
\end{lstlisting}
Using {\sf vectorized\_value\_and\_grad} now requires the {\sf vectorized\_argnums} argument to be a tuple corresponding to the $x,y$ argument indices:
\begin{lstlisting}[language=Python]
f_vvg = K.vectorized_value_and_grad(
    f, argnums=2, vectorized_argnums=(0, 1), has_aux=True
)
\end{lstlisting}
In the $f(x,y,w)$ cost function, $x,y$ are the zeroth and first arguments (which we wish to batch compute over), while $w$ is the second argument, which we wish to take derivatives with respect to.
The {\sf has\_aux} argument, if set to True, indicates that the function returns a tuple with only the first element differentiated. In our case, the first output is the loss function to be minimized, and the second auxiliary output is the predicted label {\sf yp}, which is also helpful to keep for calculating other metrics such as AUC or ROC. 
The batch computation can then be performed as follows
\begin{lstlisting}[language=Python]
X = tc.array_to_tensor([[3, 2], [1, -4], [0, 1]])
Y = tc.array_to_tensor([[0], [1], [1]])
w = tc.array_to_tensor([0.1, 0.3])
f_vvg(X, Y, w)
\end{lstlisting}
\subsubsection{Batched VQE} \label{sec:batched_vqe}
\begin{figure}[htp]
    \centering
    \includegraphics[width=0.5\textwidth]{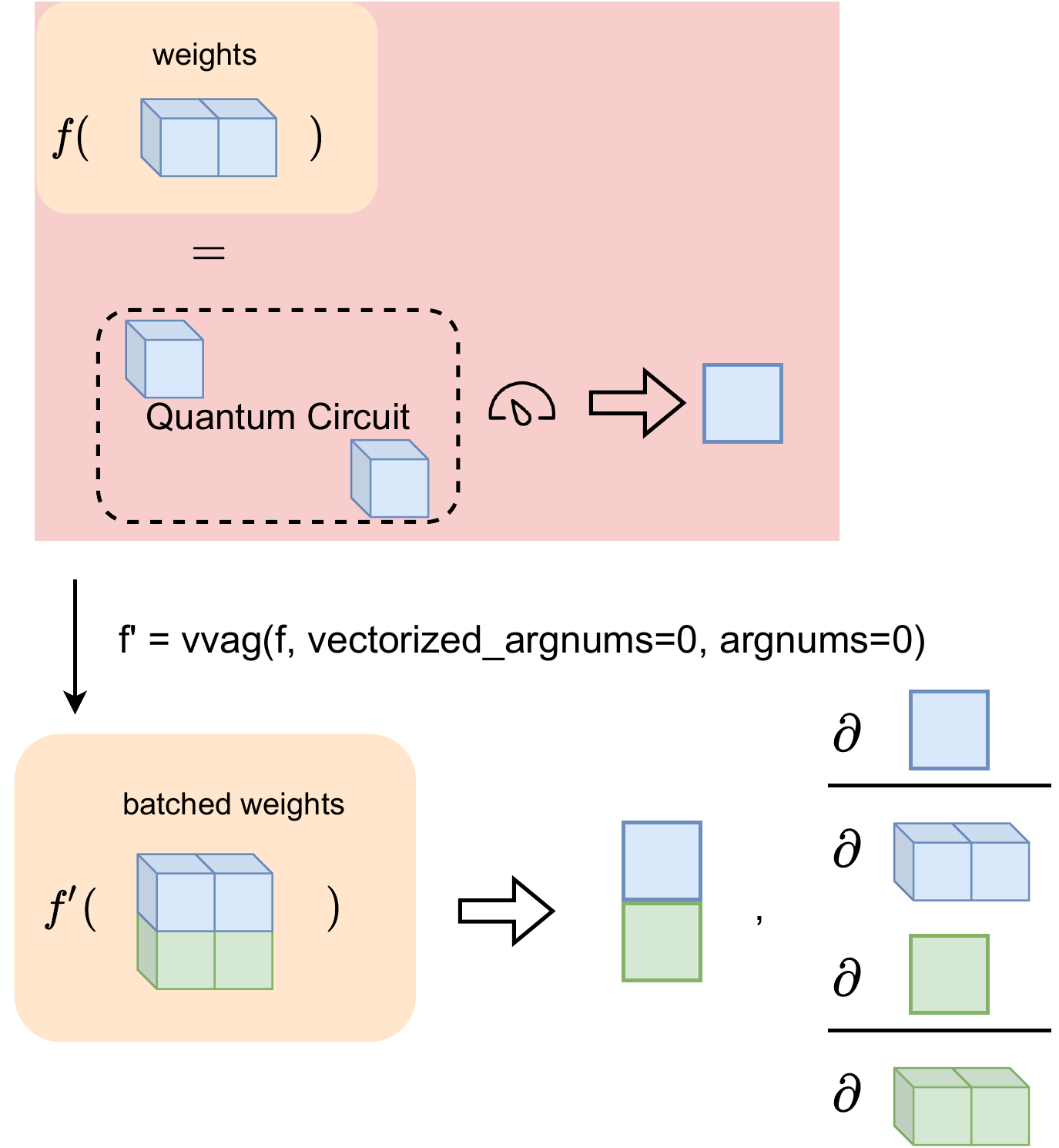}

\caption{Applying {\sf vectorized\_value\_and\_grad} for VQE with batched circuit weights. This enables multiple optimization loop evaluations to be performed at the same time. In the figure, the function f' at the bottom is transformed from the original function f defined at the top via {\sf f' = K.vectorized\_value\_and\_grad(f)}. The {\sf weights} input is both differentiated and vectorized at the same time.}
    \label{fig:batchvqe}
\end{figure}

Consider a cost function defined by a simple parameterized quantum ciruit, e.g.,
\begin{lstlisting}[language=Python]
def f(w):
    c = tc.Circuit(2)
    c.rx(0, theta=w[0])
    c.cnot(0, 1)
    c.rx(1, theta=w[1])
    e = c.expectation_ps(z=[0, 1])
    return K.real(e)
\end{lstlisting}
The function value and gradient (with respect to the weights $w[0],w[1]$) can be batch computed for multiple weights simultaneously as follows:
\begin{lstlisting}[language=Python]
f_vvag = K.vectorized_value_and_grad(f, argnums=0, vectorized_argnums=0)

W = tc.array_to_tensor([[0.1, 0.2], [0.3, 0.4]])
f_vvag(W)
\end{lstlisting}
While the above is a toy problem for illustrative purposes, combining this method with {\sf jit} can be a powerful approach to finding ground state energies via VQE, starting from multiple initial points in parameter space and finding the best local minimum reached by gradient descent.  The batched VQE workflow, which admits independent optimization loops running simultaneously, is sketched in Figure~\ref{fig:batchvqe}. 

\subsubsection{Batched Monte Carlo trajectory noise simulation}

As introduced in Section~\ref{sec:vmap_MC}, by using external random number arrays, we can simulate quantum noise via multiple Monte Carlo trajectories computed in a vectorized parallel fashion. \markedup{Each trajectory is a specific instance of the noisy circuit, where each quantum channel is replaced by a stochastically chosen operator determined by external random input.} The following example shows how to vmap quantum noise (external randomness) in \tc in a variational quantum algorithm.

\begin{lstlisting}[language=python]
nwires = 6


def f(weights, status):
    c = tc.Circuit(nwires)
    # omit the details of constructing a PQC with `weights` parameters
    for i in range(nwires):
        c.depolarizing(i, px=0.2, py=0.2, pz=0.2, status=status[i])
        # quantum noise controlled by external random number `status` argument
    loss = c.expectation_ps(x=[nwires // 2])
    loss = tc.backend.real(loss)
    return loss


# get the circuit gradient while vmapping the depolarizing noise
f_vg = tc.backend.jit(tc.backend.vvag(f, argnums=0, vectorized_argnums=1))

# random number with batch dimension
status = tc.backend.implicit_randu(shape=[batch, nwires])
f_vg(weights, status)
\end{lstlisting}

\subsection{{\sf QuOperator} and {\sf QuVector}}\label{sub:mps}

\begin{mdframed}
\textbf{Jupyter notebook: }
\fancylink{\rooturl docs/source/whitepaper/6-4-quoperator.ipynb}{6-4-quoperator.ipynb}
\end{mdframed}

{\sf tc.quantum.QuOperator}, {\sf tc.quantum.QuVector} and {\sf tc.quantum.QuAdjointVector} are data classes which behave like matrices and vectors (columns or rows) when interacting with other ingredients, while their inner structures correspond to tensor networks for efficiency and compactness.

Typical tensor network structures for a QuOperator/QuVector correspond to Matrix Product Operators (MPO) / Matrix Product States (MPS). The former represents a matrix as:
\eq{M_{i1,i2,...in; \; j1, j2,... jn}=\prod_k {T_k}^{i_k, j_k},}
i.e., a product of $d\times d$ matrices $T_k^{i_k, j_k}$, where $d$ is known as the bond dimension. Similarly, an MPS represents a vector as:
\eq{V_{i_1,...i_n} = \prod_k T_k^{i_k},}
where the $T_k^{i_k}$ are, again, $d\times d$ matrices. MPS and MPO often occur in computational quantum physics contexts, as they give compact representations for certain types of quantum states and operators. For an introductory review on MPS/MPO in quantum physics, please refer to~\cite{dmrgmps}.

{\sf QuOperator}/{\sf QuVector} objects can represent any MPO/MPS, but they can additionally express more flexible tensor network structures. Indeed, any tensor network with two sets of dangling edges of the same dimension (i.e., for each $k$, the set $\{T_k^{i_k,j_k}\}_{i_k,j_k}$ of matrices has $i_k$ and $j_k$ running over the same index set) can be treated as a {\sf QuOperator}. A general {\sf QuVector} is even more flexible, in that the dangling edge dimensions can be chosen freely; thus, arbitrary tensor products of vectors can be represented.

\begin{figure}[htp]
    \centering
    \includegraphics[width=0.6\textwidth]{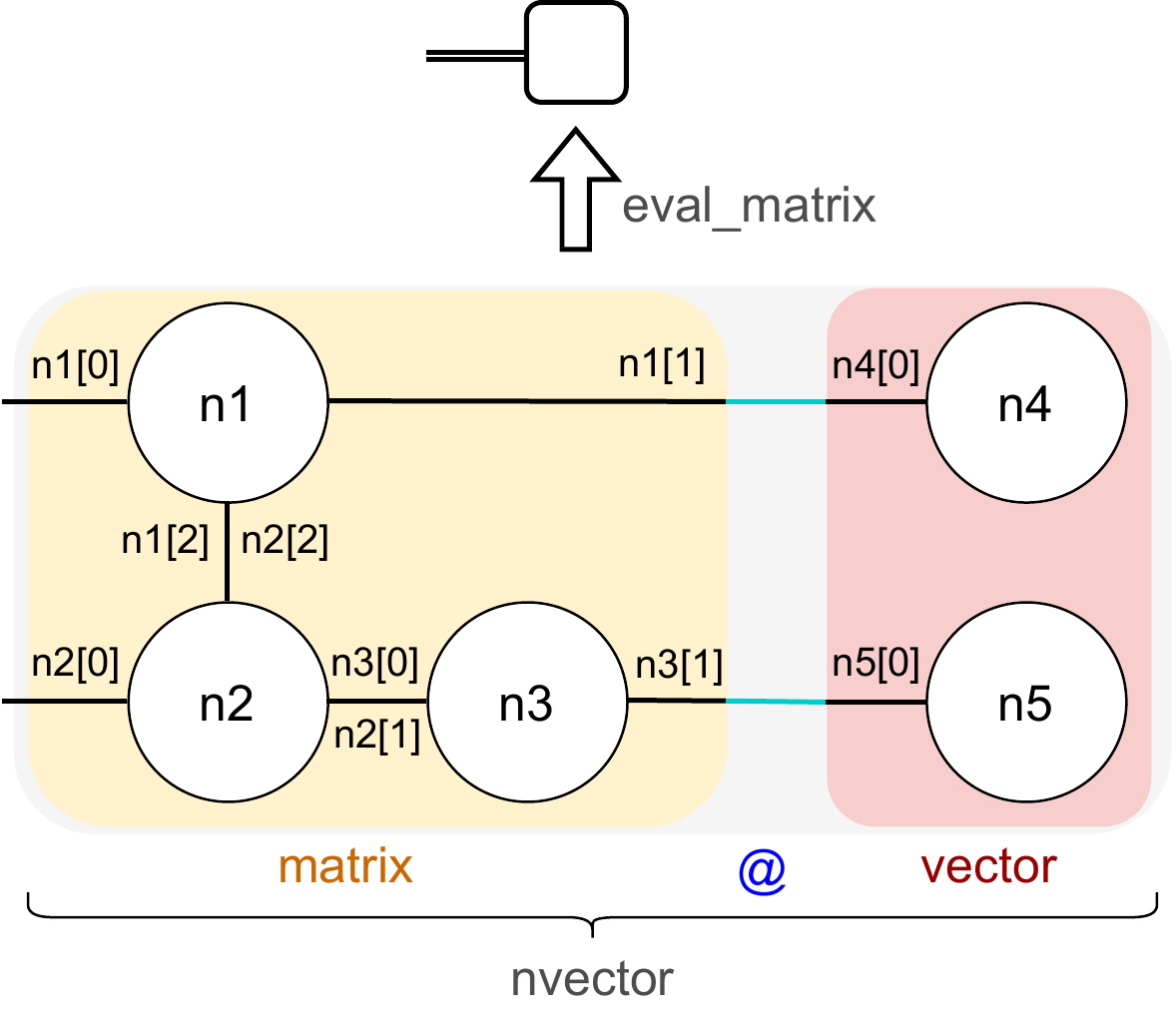}

\caption{Tensor network schematic demonstrating the usage of {\sf QuOperator} and {\sf QuVector}. \markedup{Each node is constructed from a tensor and each  {\sf QuOperator} is built by specifying the dangling edges of nodes. From the users' perspective these objects behave as matrices and vectors, with the tensor network engine responsible for maintaining their inner structures and performing calculations.} }
    \label{fig:quop}
\end{figure}

In this section, we will illustrate the efficiency and compactness of  such data structures, and show how they can be integrated seamlessly into quantum circuit simulation tasks. First, consider the following code and tensor diagram (Figure~\ref{fig:quop}) as an introduction to these data class abstractions.

\begin{lstlisting}[language=Python]
n1 = tc.gates.Gate(np.ones([2, 2, 2]))
n2 = tc.gates.Gate(np.ones([2, 2, 2]))
n3 = tc.gates.Gate(np.ones([2, 2]))
n1[2] ^ n2[2]
n2[1] ^ n3[0]

matrix = tc.quantum.QuOperator(out_edges=[n1[0], n2[0]], in_edges=[n1[1], n3[1]])

n4 = tc.gates.Gate(np.ones([2]))
n5 = tc.gates.Gate(np.ones([2]))

vector = tc.quantum.QuVector([n4[0], n5[0]])

nvector = matrix @ vector  # matrix-vector multiplication

assert type(nvector) == tc.quantum.QuVector
nvector.eval_matrix()
# array([[16.], [16.], [16.], [16.]])
\end{lstlisting}

Note that the convention in defining a {\sf QuOperator} is to first state {\sf out\_edges} (left index or row index of the matrix) and then state {\sf in\_edges} (right index or column index of the matrix). Also note that {\sf tc.gates.Gate} is just a wrapper for the {\sf Node} object in the TensorNetwork package.

As seen above, {\sf QuOperator}/{\sf QuVector} objects support matrix-matrix or matrix-vector multiplication via the {\sf{@}} operator.  Other common matrix/vector operations are also supported: 
\begin{lstlisting}[language=Python]
matrix.adjoint()            # adjoint i.e., conjugate transpose
5 * vector                  # scalar multiplicatoin
vector | vector             # tensor product
matrix.partial_trace([0])   # partial trace (of subsystem 0)
\end{lstlisting}

Matrix elements of these objects can be extracted via {\sf .eval()} or {\sf .eval\_matrix()}. The former keeps the shape information of the tensor network while the latter gives the matrix representation (i.e, as a rank 2 tensor).

\subsubsection{{\sf QuVector} as the input state for the circuit}

Since a {\sf QuVector} behaves like a regular vector, albeit with a more compact representation, it can be used as the input state to a quantum circuit instead of a state represented as a regular array. The benefit of doing so is memory efficiency. For an $n$-qubit circuit, regular vector inputs require $2^n$ complex values to be stored in memory. On the other hand, for an MPS with bond dimension $d$ represented by a {\sf QuVector}, only $O(nd^2)$ complex elements in total need to be stored.  Such compact MPS representations can be obtained, for instance, by DMRG~\cite{PhysRevLett.69.2863} calculations, and a DMRG ground state to quantum machine learning model pipeline can thus be built in \tc with the help of this feature.

The following example shows how we input the $\vert 111\rangle$ state, encoded as an MPS, to a quantum circuit. Note how {\sf mps\_inputs} argument is used when constructing the circuit.  

\begin{lstlisting}[language=python]
n = 3
nodes = [tc.gates.Gate(np.array([0.0, 1.0])) for _ in range(n)]
mps = tc.quantum.QuVector([nd[0] for nd in nodes])
c = tc.Circuit(n, mps_inputs=mps)
c.x(0)
c.expectation_ps(z=[0])  # 1
\end{lstlisting}

\subsubsection{{\sf QuVector} as the output state of the circuit}

For a given input state, a {\sf tc.Circuit} object can itself be treated as a tensor network with one set of dangling edges corresponding to the output state, and thus the whole circuit object can be regarded as a {\sf QuVector}, obtained via {\sf c.quvector()}. We can then further manipulate the circuit using {\sf QuOperator} objects.

\subsubsection{ {\sf QuOperator} as the operator to be measured}

As shown in Section~\ref{sub:pss}, Hamiltonians can also be represented by a {\sf QuOperator}. This can be a powerful and efficient approach for computing expectation values for some lattice model Hamiltonians like the Heisenberg model or transverse field Ising model (TFIM), as the MPO  form of the Hamiltonian for such short-ranged spin models has a very low bond dimension (e.g., $d=3$ for TFIM). For comparison, for an $n$-qubit TFIM Hamiltonian, the dense matrix representation stores $O(2^{2n})$ complex elements, the sparse matrix representation stores $O(n2^n)$ complex elements, while the MPO or {\sf QuOperator} representation stores only $18n$ complex elements, which scales linearly with the system size.

Here we show a toy example of measuring the expectation value of an operator represented as a {\sf QuOperator}. The quantity of interest here is $\langle Z_0 Z_1\rangle$, where the expectation is with respect to the output of a simple circuit which consists of an $X$ gate on one of the qubits. Instead of using the {\sf c.expectation()} API, we use {\sf mpo\_expectation}.

\begin{lstlisting}[language=python]
z0, z1 = tc.gates.z(), tc.gates.z()
mpo = tc.quantum.QuOperator([z0[0], z1[0]], [z0[1], z1[1]])
c = tc.Circuit(2)
c.X(0)
tc.templates.measurements.mpo_expectation(c, mpo)  # -1
\end{lstlisting}

\subsubsection{{\sf QuOperator} as a quantum gate}

Since quantum gates correspond to unitary matrices, an MPO representation for these matrices may allow for significant space efficiencies in some scenarios. A typical example is the multi-controlled gate, which admits a very compact MPO representation that can be characterized by a {\sf QuOperator} in \tc. For an $n$-qubit gate with $n-1$ control qubits, the matrix representation for this gate has $2^{2n}$ elements while the MPO representation can be reduced to bond dimension $d=2$, leading to only $16n$ elements in memory.

\tc has a built-in way to generate these efficient multi-controlled gates:
\begin{lstlisting}[language=python]
# CCNOT gate
c = tc.Circuit(3)
c.multicontrol(0, 2, 1, ctrl=[1, 0], unitary=tc.gates.x()) #if q0=1 and q2=0, apply X to q1
\end{lstlisting}
The {\sf 0,2,1} arguments refer to the qubits the gate is applied on (the ordering matters, with the final index referring to the target qubit), the {\sf unitary} argument defines the operation that is applied if all controls are activated and the {\sf ctrl} argument refers to whether the control is activated when the corresponding control qubit is in the 0 state or 1 state.   
General MPO gates expressed as a {\sf QuOperator} can also be applied via the {\sf c.mpo(*index, mpo=)} API, in the same way that general unitary matrices can be applied via the {\sf c.unitary(*index, unitary=)} API (see Section~\ref{sec:basics}).

\subsection{Custom contraction settings}

\begin{mdframed}
\textbf{Jupyter notebook: }
\fancylink{\rooturl docs/source/whitepaper/6-5-custom-contraction.ipynb}{6-5-custom-contraction.ipynb}
\end{mdframed}

By default, \tc uses a greedy tensor contraction path finder provided by the opt\_einsum package. While this is typically satisfactory for moderately sized quantum circuits, for circuits with 16 qubits or more, we recommend using customized contraction path finders provided by the user or third-party packages.

A simple quantum circuit can be constructed using the following code as a testbed for different contraction methods:
\begin{lstlisting}[language=Python]
def testbed():
    n = 40
    d = 6
    param = K.ones([2 * d, n])
    c = tc.Circuit(n)
    c = tc.templates.blocks.example_block(c, param, nlayers=d, is_split=True)
    # the two qubit gate is split and truncated via SVD decomposition
    return c.expectation_ps(z=[n // 2], reuse=False)
    # by reuse=False, we compute the expectation as a single tensor network instead of first computing the wavefunction

\end{lstlisting}
By using {\sf tc.templates.blocks.example\_block}, a circuit with $d$ layers of $\exp(i\theta ZZ)$ gates and $R_x$ gates is created. When {\sf is\_split} is {\sf True}, each two-qubit gate will not be treated as an individual tensor but will be split into two connected tensors via singular value decomposition (SVD), which further simplifies the tensor network structure of the corresponding circuit. 
The task is to calculate the expectation value of the Z operator on the middle (i.e, $n/2$-th) qubit.

The API for contraction setup is {\sf tc.set\_contractor}. 
In our example, $2n\times d$ tensors need to be contracted since single-qubit gates can be absorbed into two-qubit gates when {\sf preprocessing} is set to {\sf True} in {\sf set\_contractor}.
We have some built-in contraction path finder options such as "greedy", "branch", and "optimal" from opt-einsum \cite{Smith2018}, though only the default "greedy" option is suitable for circuit simulation tasks as other options require time exponential in the number of qubits. 
The {\sf contraction\_info} option in this setup API, if set True, will print the contraction path information after the path searching. Metrics that measure the quality of a contraction path include 
\begin{itemize}
    \item \textbf{FLOPs}: the total number of computational operations required for all matrix multiplications involved when contracting the tensor network via the given path. This metric characterizes the total simulation time.
    \item \textbf{WRITE}: the total size (the number of elements) of all tensors -- including intermediate tensors -- computed during the contraction. 
    \item \textbf{SIZE}: the size of the largest intermediate tensor stored in memory.
\end{itemize}
Since simulations in \tc are AD-enabled, where all intermediate results need to be cached and traced, the more relevant spatial cost metric is write instead of size.

\subsubsection{Customized contraction path finder}
For large quantum circuits, the performance of the default "greedy" contraction path finder may not be satisfactory. If this is the case, a custom contraction path finder can be used to enhance the performance of contraction by finding better paths in terms of flops (time) and writes (space). Here we use the path finder provided by the third-party cotengra package , a python library for contracting tensor networks or computing einsum expressions. The way to use the cotengra path finder in \tc is as follows:
\begin{lstlisting}[language=Python]
import cotengra as ctg

opt = ctg.ReusableHyperOptimizer(
    methods=["greedy", "kahypar"],
    parallel=True,
    minimize="write",
    max_time=120,
    max_repeats=1024,
    progbar=True,
)
tc.set_contractor("custom", optimizer=opt, preprocessing=True, contraction_info=True)
testbed()
\end{lstlisting}
A number of parameters are used to configure a contraction path finder {\sf opt} in cotengra: {\sf method} decides the strategy this path finder will be based on. {\sf minimize} decides the score function you want to minimize during the path finding, and can be set as "write", "flops", "size" or a combination of these. A time limit and a limit on the number of trial contraction trees can also be set using {\sf max\_time} and {\sf max\_repeats} respectively. For more details, refer to the cotengra documentation. You can also design your own contraction path finder as long as you provide an {\sf opt} function compatible with the interface of the {\sf opt\_einsum} optimizer. 

\subsubsection{Subtree reconfiguration}\label{sub:reconf}
Given a contraction path, e.g. given by a "greedy" search, its performance can be further enhanced by conducting a so-called subtree reconfiguration. This process repeatedly optimizes subtrees of the whole contraction tree, and in practice often results in a better contraction path. This can be done in \tc as follows:
\begin{lstlisting}[language= Python]
opt = ctg.ReusableHyperOptimizer(
    minimize="combo",
    max_repeats=1024,
    max_time=120,
    progbar=True,
)


def opt_reconf(inputs, output, size, **kws):
    tree = opt.search(inputs, output, size)
    tree_r = tree.subtree_reconfigure_forest(
        progbar=True, num_trees=10, num_restarts=20, subtree_weight_what=("size",)
    )
    return tree_r.get_path()


tc.set_contractor(
    "custom",
    optimizer=opt_reconf,
    contraction_info=True,
    preprocessing=True,
)


testbed()
\end{lstlisting}
Notice that {\sf subtree\_reconfigure\_forest} is used after finding a contraction tree. In this function, you can set the number of trees in the random forest whose contraction paths will be updated, and also the metric to be optimized in the subtrees. In the above example, a user customized function {\sf opt\_reconf} is fed into contractor setup as a legal contraction path finder.

As mentioned earlier, there are three metrics to measure the quality of a contraction path. By setting different score functions (changing the {\sf minimize} parameter), the resulting contraction path given by these contractors will exhibit different properties. Table~\ref{tab:contractor} summarizes the contraction performance of different contraction strategies for our example case ($n=40, d=6$). As we can see, the cotengra optimizer and subtree reconfiguration can greatly improve the quality of the contraction path and improve the efficiency of quantum circuit simulation. For example, we gain more than a factor of two improvement in simulation time and simulation space compared to the default contractor, and the degree of improvement can increase for larger system sizes.

\begin{table}\centering
\begin{tabular}{c|c|c|c|c}
contractor    & reconfiguration  & log10[FLOPs] & log2[SIZE] & log2[WRITE] \\ 
     \hline 
default &    &   7.373    &    12   &   20.171   \\
cotengra({\sf "flops"}) &   &   7.080    &  13 & 20.493     \\
cotengra({\sf "flops"}) & subtree({\sf "flops"})  &   7.006    &   12    &   20.069   \\
cotengra({\sf "flops"}) & subtree({\sf "size"})  &   7.006    &   12    &   20.075   \\
cotengra({\sf "write"}) &   &  7.442     &    14   &   19.061   \\
cotengra({\sf "write"}) &subtree({\sf "flops"})  &  7.000     &    12   &    19.988  \\
cotengra({\sf "write"}) &subtree({\sf "size"})  &  7.017     &    12   &    19.958  \\
cotengra({\sf "combo"}) &   &  7.480     &    14   &   19.061   \\
cotengra({\sf "combo"}) &subtree({\sf "flops"})  &  7.003     &    12   &    20.000  \\
cotengra({\sf "combo"}) &subtree({\sf "size"})  &  7.011    &    12   &    19.885  \\
     
\end{tabular}
\caption{Performance of different contractor settings which include the default {\sf opt\_einsum} contractor and cotengra contractors with and without subtree reconfiguration. Parameters in the brackets indicate the score function used during the path searching and reconfiguration. For cotengra contractors, we set {\sf max\_repeats=1024, max\_time=120} and {\sf method=["greedy","kahypar"]}. "combo" means the score function is a combination of "flops" and "write" ({\sf flops+64$\times$ write} by default). For subtree reconfiguration, we set {\sf num\_trees =20, num\_restarts=20}. Results shown are from one run, and performance may vary from run to run since these algorithms are intrinsically random.  }\label{tab:contractor}
\end{table}

\subsection{Advanced automatic differentiation} \label{sec:adqiskit}

\begin{mdframed}
\textbf{Jupyter notebook: \fancylink{\rooturl docs/source/whitepaper/6-6-advanced-automatic-differentiation.ipynb}{6-6-advanced-automatic-differentiation.ipynb}}

\end{mdframed}

\tc provides backend-agnostic wrappers to a number of advanced AD features, useful in a variety of quantum circuit simulation scenarios.  In the remainder of this section we will illustrate these using the following circuit example:

\begin{lstlisting}[language=Python]
n = 6
nlayers = 3


def ansatz(thetas):
    c = tc.Circuit(n)
    for j in range(nlayers):
        for i in range(n):
            c.rx(i, theta=thetas[j])
        for i in range(n - 1):
            c.cnot(i, i + 1)
    return c


def psi(thetas):
    c = ansatz(thetas)
    return c.state()
\end{lstlisting}

\bp{Jacobian ({\sf jacfwd} and {\sf jacrev}).} 
Given an $n$-input, $m$-output function $f$ the $n\times m$  Jacobian matrix is given by
\eq{
J_f := \frac{\pd f}{\pd x} =\begin{pmatrix} \frac{\pd f_1}{\pd x_1} & \ldots & \frac{\pd f_1}{\pd x_n} \\ 
\vdots & \ddots \\  
\frac{\pd f_m}{\pd x_1} & \ldots & \frac{\pd f_m}{\pd x_n}\end{pmatrix}
}
By the chain rule, the Jacobian matrix of a composition of functions is the product of the Jacobians of the composed functions (evaluated at appropriate points). e.g. if $h:\mb{R}^n\ra\mb{R}^p$, $g:\mb{R}^p\ra\mb{R}^q$, $f:\mb{R}^q\ra\mb{R}^m$ and $y:\mb{R}^n\ra \mb{R}^m$ with $y(x) = f(g(h(x)))$ then
\eq{
\frac{\pd y}{\pd x} &= \frac{\pd f(b)}{\pd b}\cdot \frac{\pd g(a)}{\pd a}\cdot \frac{\pd h(x)}{\pd x}\\
&= J_f(b) \cdot J_g(a) \cdot J_h(x)
}
where $a = h(x), b= g(a)$ and $\cdot$ denotes matrix multiplication.  Forward mode AD and reverse mode AD (`backpropagation') are two approaches to computing a composite Jacobian, and differ in the order in which the products are computed.  Forward mode AD computes the above product from right to left, i.e. $J_y = J_f(b)\cdot \lp J_g(a) \cdot J_h(x)\rp$ at a cost of $pqn + qnm$ multiplications. Taking $f$ to be the output state $\psi(\theta)$ of the above circuit, this is computed as
\begin{lstlisting}[language=Python]
thetas = K.implicit_randn([nlayers])
jac_fwd_function = K.jacfw(psi)
jac_fw = jac_fwd_function(thetas)
\end{lstlisting}
Reverse mode AD computes the product from left to right, i.e. $J_y = \lp J_f(b)\cdot \rp J_g(a) \cdot J_h(x)$ at a cost of $mpq + pnm$ multiplications:
\begin{lstlisting}[language=Python]
jac_rev_function = K.jacrev(psi)
jac_rev = jac_rev_function(thetas)
\end{lstlisting}
The relative efficiency of these methods depends on the input and output dimensions of the functions involved. For instance, when $p=q$ forward mode AD is advantageous if $n \ll m$ (i.e., the input dimension is much smaller than the output dimension, corresponding to a `tall' Jacobian) and vice versa for reverse mode AD.

\bp{Jacobian-vector product (jvp).} Computing the product of the Jacobian with a vector $v$ (i.e. the directional derivative) can be a useful primitive as it utilizes forward mode AD and is suitable when the output dimension is much larger than the input. For instance, setting $v = e_i$ (i.e. the vector with a $1$ in the $i$-th coordinate and zeroes elsewhere) gives the vector of partial derivatives
\eq{
J_f e_i = \begin{pmatrix}\frac{\pd f_1}{\pd x_i}& ,\ldots, &
\frac{\pd f_m}{\pd x_i}
\end{pmatrix}^\top
}
Taking $v = (1.0,0,0)$, the value of $\psi$ and the Jacobian-vector product $\frac{\partial \psi}{\partial \theta_0}$ can be evaluated (e.g., at the point $\theta=(0.1,0.2,0.3)$)  as follows:
\begin{lstlisting}[language=Python]
state, partial_psi_partial_theta0 = K.jvp(
    psi,
    tc.array_to_tensor([0.1, 0.2, 0.3]),
    tc.array_to_tensor([1.0, 0, 0], dtype="float32"),
)
\end{lstlisting}

\bp{Quantum Fisher Information ({\sf qng}).} The Quantum Fisher Information (QFI) is an important concept in quantum information, and can be utilized in so-called quantum natural gradient descent optimization~\cite{Stokes_2020} as well as variational quantum dynamics \cite{yuan2019theory, PhysRevLett.125.010501}. 

There are several variants of QFI-like quantities, all of which depend on the evaluation of terms of the form $\langle \partial_i \psi \vert \partial_j \psi\rangle - \langle \partial_i \psi \vert \psi\rangle\langle \psi \vert \partial_j \psi\rangle$. Such quantities are easily obtained with advanced AD frameworks, by first computing the Jacobian for the output state and then vmapping the inner product over Jacobian rows. The detailed efficient implementation can be found at the codebase. Here we directly call the corresponding API to obtain the quantum natural gradient.
\begin{lstlisting}[language=python]
from tensorcircuit.experimental import qng

# function to get qfi given circuit parameters
qfi_fun = K.jit(qng(psi))

# suppose the vanilla circuit gradient is `grad`
# then we can obtain quantum natural gradient as
ngrad = tc.backend.solve(qfi_fun(thetas), grad, assume_a="sym")
\end{lstlisting}

\bp{Hessian ({\sf hessian}).} The Hessian  $H_{ij} = \frac{\partial \langle H\rangle_{\theta}}{\partial  \theta_i \partial \theta_j}$ of a parameterized quantum circuit can be computed as (taking $H = Z_0$ for simplicity)
\begin{lstlisting}[language=Python]
def h(thetas):
    c = ansatz(thetas)
    return c.expectation_ps(z=[0])

# hess is the Hessian function which takes thetas as input
hess = K.hessian(h)
\end{lstlisting}
which can  then also be jitted to make multiple evaluations more efficient:
\begin{lstlisting}[language=Python]
hess_jit = K.jit(hess)
\end{lstlisting}
Information on the Hessian matrix may be useful in investigating loss landscapes,  or for second order optimization routines.

\begin{figure}[htp]
    \centering
    \includegraphics[width=0.7\textwidth]{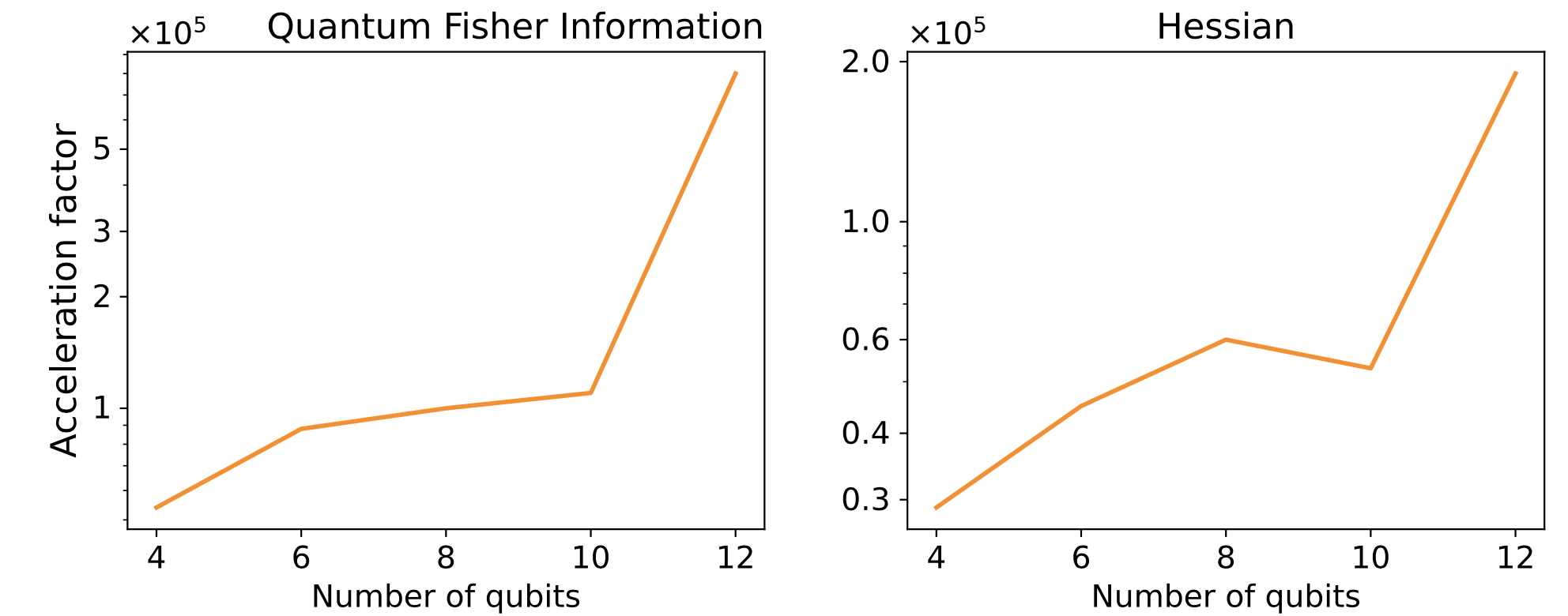}

\caption{Acceleration factor for \tc over Qiskit for the evaluation of QFI and Hessian information. For 4 and 6 qubit systems, we use PQC comprising two blocks of CNOT+Rx+Rz+Rx gates, while for the larger systems we use four such blocks. The simulation runs on AMD EPYC 7K62 CPU 2.60GHz, and \tc results use the JAX backend. As the Qiskit running times are long (e.g., the 12-qubit QFI calculation takes more than $20{,}000$s $\sim 5.5$ hours) the acceleration factors presented above are based on only a single Qiskit evaluation. In contrast, the \tc running times are low enough ($0.026$s for 12-qubit QFI) that we average over multiple runs.}
    \label{fig:gradient}
\end{figure}

\bp{\bra{\psi} H \ket{\partial \psi}.}  In variational quantum dynamics problems (see, e.g.~\cite{yuan2019theory}) one often wishes to compute quantities of the form
\eq{
\bra{\psi(\theta)} H
\frac{\partial \ket{\psi(\theta)}}{\partial \theta_i}.
}
This can be done via the {\sf stop\_gradient} API, which prevents certain parameters from being differentiated. Again, taking $H=Z_0$, we can define an appropriate function for which only the parameters corresponding to $\ket{\psi(\theta)}$ will be differentiated:
\begin{lstlisting}[language=Python]
z0 = tc.quantum.PauliStringSum2Dense([[3, 0, 0, 0, 0, 0]])


def h(thetas):
    w = psi(thetas)
    w_left = K.conj(w)
    w_right = K.stop_gradient(w)
    w_left = K.reshape([1, -1])
    w_right = K.reshape([-1, 1])
    e = w_left @ z0 @ w_right
    return K.real(e)[0, 0]
\end{lstlisting}
Then, gradients can be computed as usual:
\begin{lstlisting}[language=Python]
psi_h_partial_psi = K.grad(h)
\end{lstlisting}

With the advanced automatic differentiation infrastructure, we can obtain quantum circuit gradient related quantities, such as those listed above, much more quickly than via traditional quantum software that utilizes parameter shifts to evaluate gradients. In Figure~\ref{fig:gradient}, we show the acceleration of QFI and Hessian computations compared to Qiskit. The benchmark code is detailed in \fancylink{\rooturl/examples/gradient_benchmark.py}{ examples/gradient\_benchmark.py}. From the data, we see that for even moderate-sized quantum circuits, \tc can achieve a speedup over Qiskit of nearly a {\it million times}.

\section{Integrated examples}\label{sec:examples_benchmarks}

\subsection{Molecular VQE with \tc and OpenFermion}\label{subsec:h2ovqe}

\begin{mdframed}
\textbf{Jupyter notebook: }
\fancylink{\rooturl docs/source/tutorials/vqe_h2o.ipynb}{tutorials/vqe\_h2o.ipynb}
\end{mdframed}

{\bp {Background}}. Quantum computing is envisioned to be a powerful tool for quantum simulation and quantum chemistry tasks~\cite{RevModPhys.92.015003, Cao2019}. In the {\it highlighted features} below we show how to interface \tc with OpenFermion~\cite{openfermion} to compute the ground state energy of an H$_2$O molecule, while in the {\it benchmarking} section we compare the performance of \tc with other software for the VQE energy computation of a transverse field Ising model.

{\bp {Highlighted features}}. OpenFermion is an open-source Python package that provides an efficient interface between quantum chemistry and quantum computing. In particular, it provides a convenient method for generating molecular Hamiltonians and converting them into qubit Hamiltonians compatible with simulating in quantum circuits. To make use of this, we provide the {\sf tc.templates.chems.get\_ps} API, which provides an interface between \tc and OpenFermion, and converts the OpenFermion qubit Hamiltonian object into the Pauli structures and weights tensors used by \tc (see Section~\ref{sec:structures}). The relevant code snippet used to generate the Hamiltonian representation in \tc via OpenFermion is shown below.

\begin{lstlisting}[language=python]
from openfermion.chem import MolecularData, geometry_from_pubchem
from openfermion.transforms import get_fermion_operator, jordan_wigner
from openfermionpyscf import run_pyscf

multiplicity = 1
basis = "sto-3g"
# 14 spin orbitals for H2O
geometry = geometry_from_pubchem("h2o")
molecule = MolecularData(geometry, basis, multiplicity)
# obtain H2O molecule object
molecule = run_pyscf(molecule, run_mp2=True, run_cisd=True, run_ccsd=True, run_fci=True)
print(molecule.fci_energy, molecule.ccsd_energy, molecule.hf_energy)

mh = molecule.get_molecular_hamiltonian()
# get fermionic Hamiltonian
fh = get_fermion_operator(mh)
# get qubit Hamiltonian via Jordan-Wigner transformation
jw = jordan_wigner(fh)
# converting to Pauli structures in tc
structures, weights = tc.templates.chems.get_ps(jw, 14)
# build sparse numpy matrix representation for the Hamiltonian
ma = tc.quantum.PauliStringSum2COO_numpy(strutcures, weights)
\end{lstlisting}

\bp{Benchmarking}. We provide some benchmark data for TFIM VQE evaluation here. Specifically, we compute the TFIM energy expectation value and corresponding circuit gradients (with respect to the circuit parameters). With \tc, we use the (potentially less efficient) explicit {\sf for} loop to perform the Pauli string summation (see Section~\ref{sub:explicitloop}) to obtain a fair comparison, with our benchmarking focus on the efficiency of evaluating the PQC and its gradients. 
See \fancylink{\rooturl benchmarks}{benchmarks} for benchmark setup and details.
Results are summarized in Table~\ref{tab:tfimvqebm}. \markedup{Note that while we consider the TFIM model in our benchmark tests, this only differs from molecular systems in the number and type of Pauli strings that appear in the Hamiltonian.}

\markedup{We do not perform similar benchmark tests with other common quantum simulator packages such as Qiskit, Cirq and ProjectQ, as these do not support automatic differentiation, relying instead on the parameter shift technique to evaluate gradients. As the computational complexity of this approach scales linearly with the number of variational parameters, this is inefficient for circuits with a large number of parameters. For example, using Qiskit with its built-in parameter-shift gradient framework, the task in Table~\ref{tab:tfimvqebm} (with $n=10$, $d=3$ takes $125$s per iteration -- more than $10^5$ times slower than \tc. Similar results have already been presented in Section~\ref{sec:adqiskit}.  }
%Therefore, we don't compare with quantum simulators of without AD support as we focus on NISQ scenarios where circuit gradients play a central role. 
%We also remark that with the latest version of cuquantum, Pennylane, Pennylane-Lightning[gpu] and the official example, we see no AD support (adjoint method) on lightning.gpu backend in the numerical experiments.

\begin{table}\centering
\begin{tabular}{c|c|c|c}
    PQC  &~ $n=10$, $d=3$ ~&  ~$n=16$, $d=16$~ &  ~$n=22$, $d=11$~\\ 
     \hline 
   \markedup{Pennylane (CPU)} & \markedup{0.012} & \markedup{0.31} &  \markedup{24.84} \\
Pennylane (GPU) &   0.067    &    0.68   &  OOM\\
TensorFlow Quantum & 0.005 & 0.026 & 0.68 \\
Qibo (GPU) & 0.033 & 0.198 & OOM  \\
\textbf{\tc (CPU)} & 0.00077 & 0.078 & 4.70 \\
\textbf{\tc (GPU)}& 0.0026 & 0.023 & 0.19

\end{tabular}
\caption{Performance benchmarks (running time in seconds) for value and gradient evaluations of $n$-qubit, $d$-layer parameterized quantum circuits with a \markedup{one-dimensional open boundary condition} TFIM Hamiltonian objective function. \markedup{Each layer of the hardware efficient ansatz comprises $n-1$ Rzz gates arranged in a cascading ladder layout and one layer of Rx gates.} \tc results use the JAX backend. GPU simulations use the Nvidia V100 32G GPU while CPU simulations use Intel(R) Xeon(R) Platinum 8255C CPU @ 2.50GHz.  OOM indicates that the GPU memory is insufficient to run the corresponding benchmark code. Note that TensorFlow Quantum only currently supports quantum circuit simulations on CPU. \markedup{Qibo results use the TensorFlow backend since this is the only Qibo backend that supports automatic differentiation. Pennylane CPU results use the Pennylane-Lightning C++ backend, while the GPU results make use of vectorized parallelism and JIT via the JAX backend. At the time of writing, the GPU version of Pennylane-Lightning does not support automatic differentiation.}  }
\label{tab:tfimvqebm}
\end{table}

\subsection{Quantum machine learning}

\begin{mdframed}
	\textbf{Jupyter notebook: }
	\fancylink{\rooturl docs/source/tutorials/mnist_qml.ipynb}{tutorials/mnist\_qml.ipynb}
\end{mdframed}

{\bp {Background}}. Quantum and hybrid quantum-classical neural networks are popular approaches to NISQ era quantum computing, and both can be easily modelled and tested in \tc. In the {\it highlighted features} below, we illustrate how to build a hybrid machine learning pipeline in \tc, and in the {\it benchmarking} section we compare \tc with other quantum software for performing batched supervised learning on the MNIST dataset using a parameterized quantum circuit.

{\bp {Highlighted features}}. Seamless integration of quantum and classical neural networks can be obtained by wrapping the \tc {\sf tc.Circuit} object (with weights as input and expectation value as output) in a {\sf QuantumLayer}, which is a subclass of the Keras {\sf Layer}. The following code snippet shows how such a wrapper is implemented and used:

\begin{lstlisting}[language=python]
	def qml_ys(x, weights, nlayers):
	n = 9
	weights = tc.backend.cast(weights, "complex128")
	x = tc.backend.cast(x, "complex128")
	c = tc.Circuit(n)
	for i in range(n):
	c.rx(i, theta=x[i])
	for j in range(nlayers):
	for i in range(n - 1):
	c.cnot(i, i + 1)
	for i in range(n):
	c.rx(i, theta=weights[2 * j, i])
	c.ry(i, theta=weights[2 * j + 1, i])
	ypreds = []
	for i in range(n):
	ypred = c.expectation([tc.gates.z(), (i,)])
	ypred = tc.backend.real(ypred)
	ypred = (tc.backend.real(ypred) + 1) / 2.0
	ypreds.append(ypred)
	# return <z_i> as an n dimensional vector
	return tc.backend.stack(ypreds)
	
	
	# wrap the quantum function in a Keras layer
	ql = tc.keras.QuantumLayer(partial(qml_ys, nlayers=nlayers), [(2 * nlayers, 9)])
	# build the hybrid Keras model with quantum and classical parts
	model = tf.keras.Sequential([ql, tf.keras.layers.Dense(1, activation="sigmoid")])
	
	# train as a normal Keras model
	model.compile(
	loss=tf.keras.losses.BinaryCrossentropy(),
	optimizer=tf.keras.optimizers.Adam(0.01),
	metrics=[tf.keras.metrics.BinaryAccuracy()],
	)
	model.fit(x_train, y_train, batch_size=32, epochs=100)
\end{lstlisting}

\bp{Benchmarking}. We benchmark \tc against other software for binary classification (`3' vs.\ `6') of the MNIST dataset, using quantum machine learning with batched inputs. For software with only parameter-shift gradient support, each PQC must be evaluated $O(np)$ times, where $np$ is the number of circuit parameters. This is much slower than AD-enabled simulators, where only one evaluation of the PQC suffices to obtain all circuit weight gradients. Therefore, we only compare \tc with other AD enabled software such as TensorFlow Quantum and Pennylane (With Pennylane, we utilize its JAX backend simulator, and use jit and vmap tricks to increase performance). The \tc results use the default greedy contraction path finder, and further improvements are possible with customized contraction path finders. See \fancylink{\rooturl benchmarks}{benchmarks} for full benchmark details and Table~\ref{tab:qmlbm} for results.

From the benchmarks on the standard VQE and QML task sets in the previous two examples, we see
that \tc can indeed bring substantial speedups in quantum circuit simulation. The acceleration is more impressive on GPU, especially when the circuit size or the batch dimension is large.

\begin{table}\centering
	\begin{tabular}{c|c|c|c}
		batch size  &~ 32 ~&  ~128~ &  ~512~\\ 
		\hline 
		\markedup{Pennylane (CPU)} &   \markedup{0.58}   &    \markedup{2.21}  &  \markedup{8.80}\\
		Pennylane (GPU) &   0.042$^\ast$   &    0.0089   &   0.020\\
		TensorFlow Quantum & 0.058 & 0.24 & 0.49 \\
		\textbf{\tc (CPU)} & ~0.0070~ &~ 0.021~ & 0.085\\
		\textbf{\tc (GPU)} & 0.0035 & 0.0039 & ~0.0054 ~

	\end{tabular}
	\caption{Performance benchmarks (running time in seconds) for value and gradient evaluation of a PQC ($n=10$ qubits, circuit depth $p=3$) with classification square distance error as the objective function and batched dataset input. \tc results use the JAX backend. GPU simulations use the Nvidia V100 32G GPU while CPU simulations use Intel(R) Xeon(R) Platinum 8255C CPU @ 2.50GHz. Note that TensorFlow Quantum only runs circuit simulations on CPU.  \markedup{Pennlyane CPU results use the Pennylane-Lightning C++ backend, together with the Pennylane native batch (broadcast) paradigm. Pennylane GPU results make use of vectorized parallelism and JIT via the JAX backend.}  *The Pennylane running time on GPU is indeed higher for smaller batch size.}
	\label{tab:qmlbm}
\end{table}

\subsection{Demonstration of barren plateaus}

\begin{mdframed}
\textbf{Jupyter notebook: }
\fancylink{\rooturl docs/source/tutorials/barren_plateaus.ipynb}{tutorials/barren\_plateaus.ipynb}
\end{mdframed}

{\bp {Background}}. The so-called barren plateaus phenomenon refers to gradients of random circuits vanishing exponentially quickly as the number of qubits \markedup{or the circuit depth} increases \cite{McClean2018}.  To demonstrate this numerically requires computing the circuit gradient variance over different circuit structures (i.e., choice of random gates in the circuit) and circuit weights. Gradients can be obtained via automatic differentiation, while the different circuit weights can be vectorized and jitted to boost performance. In addition, as in this example, the different circuit structures can also be vectorized and jitted to obtain further speed up.

{\bp {Highlighted features}}.  This example showcases how {\sf jit} and {\sf vmap} can be applied to different circuit architectures. The ability to vmap circuit structures was introduced in Section~\ref{sub:vmapcircuit}. Here, we give more details on how to encode different circuit structures via a tensor parameter as input. The core part of the circuit construction is as follows.
\begin{lstlisting}[language=python]
Rx = tc.gates.rx
Ry = tc.gates.ry
Rz = tc.gates.rz

# params is a tensor for the circuit weights with shape [n_qubits, n_layers]
# seeds is a tensor for the circuit structure with shape [n_qubits, n_layers]

c = tc.Circuit(n_qubits)
for l in range(n_layers):
    for i in range(n_qubits):
        c.unitary_kraus(
            [Rx(params[i, l]), Ry(params[i, l]), Rz(params[i, l])],
            i,
            prob=[1 / 3, 1 / 3, 1 / 3],
            status=seeds[i, l],
        )
    for i in range(n_qubits - 1):
        c.cz(i, i + 1)
\end{lstlisting}

The {\sf seeds} tensor controls the circuit architecture via the {\sf unitary\_kraus} API. This API tells the circuit to stochastically attach one gate from {\sf Rx, Ry, Rz} with probability $1/3$ each. The {\sf status} argument externalizes the random number generation (see Section~\ref{sec:vmap_MC}). Namely, when {\sf status} is less than $1/3$, the first gate is applied, when {\sf status} is in the range $[1/3, 2/3]$, the second gate in the list is applied and so on. Therefore, by generating a random number array {\sf seeds = K.implicit\_randu(size=[n\_qubits, n\_layers])}, we can generate different random circuit s. The simulation over different architectures can be vectorized and jitted by generating {\sf seeds} with an extra batch dimension corresponding to the number of different architectures you wish to compute in parallel.

\begin{table}\centering
	\begin{tabular}{c|c|c|c|c}
~&	\markedup{Pennylane (CPU)} 	& TensorFlow Quantum &  \textbf{\tc (CPU)} & \textbf{\tc (GPU)} \\ 
		\hline 
		time (s)  & \markedup{155.28} &   6.24    &    0.12   & 0.011\\
		
	\end{tabular}
	\caption{Performance benchmark for gradient variance evaluation over different random circuit architectures and circuit weights (10 qubits, 10 layers, 100 different circuits). \tc results use the JAX backend. \markedup{Pennylane results use the Pennylane-Lightning backend with highly optimized C++ code.} GPU simulations use the Nvidia V100 32G GPU while CPU simulations use Intel(R) Xeon(R) Gold 6133 CPU @ 2.50GHz. Note that TensorFlow Quantum only supports CPU for quantum circuit simulation.}
	\label{tab:bpbench}
\end{table}

\bp{Benchmarking}. We benchmark the gradient variance computation over different random circuit weights and different circuit structures using TensorFlow Quantum, \markedup{Pennylane} and \tc, (see details in \newline\fancylink{\rooturl examples/bp_benchmark.py}{examples/bp\_benchmark.py}). The \markedup{gradient} computational times for 100 random circuit constructions, each a 10-qubit, 10-layer circuit,  are shown in Table~\ref{tab:bpbench}.  With $10\times 10$ circuit weights and $100$ different circuits, the combination of {\sf vmap} and {\sf jit}  provides \tc with a more than {\emph {five hundred times}} speedup over TensorFlow Quantum for this task. \markedup{As Pennylane lacks the capability to batch compute the trainable parameters and different circuit structures, the gradient evaluations must be performed sequentially by a na\"ive loop, which leads to significantly longer times even on the fastest Pennylane-Lightning backend.}

\bp{The choice of benchmark problem sets:} \markedup{The previous three subsections gave benchmark results for VQE, QML and the barren plateaus problems, respectively. These three tasks were chosen to showcase and benchmark the paradigms and features that quantum software can support. VQE problems can benefit from AD, QML problems can benefit from AD and VMAP on non-differentiable inputs, while the barren plateaus problem can be accelerated by AD as well as VMAP on differentiable inputs and on circuit structures. In all cases, JIT and GPU support (if compatible with AD and VMAP) can further increase the efficiency of the corresponding quantum simulations.}

\subsection{Very large circuit simulation}

\begin{mdframed}
\textbf{Python script: }
 \fancylink{\rooturl examples/vqe_extra_mpo.py}{examples/vqe\_extra\_mpo.py}
\end{mdframed}

{\bp {Background}}. As previously mentioned, tensor network quantum simulators do not face the same memory bottlenecks that limit full state simulators, and can thus simulate larger numbers of qubits as long as circuit connectivity and depth are reasonably low. In the {\it highlighted features} and {\it benchmarking} sections below, we consider a one dimensional TFIM VQE workflow on 600 qubits with seven layers of two-qubit gates, arranged in a cascading ladder layout, that estimates energy with more than 99\% accuracy compared to ground truth. \markedup{The Hamiltonian we consider has open boundary condition and sits at the critical point, which is believed to be the most difficult point to simulate on the phase diagram.}

\begin{figure}[htp]
    \centering
    \includegraphics[width=0.6\textwidth]{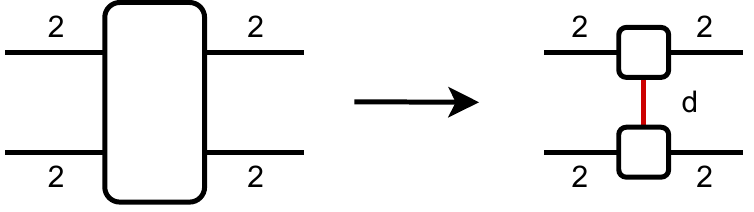}

\caption{SVD decomposition of a two-qubit gate as a tensor network. This operation can be used to reduce the complexity of the underlying network. While the bond dimension $d$ after decomposition for general two-qubit gates is 4, for certain types of two-qubit gates it can be lower. For example, in our case, where two-qubit parameterized gates are of the form $\exp{(i\theta XX)}$, $d=2$.}
    \label{fig:split}
\end{figure}

{\bp {Highlighted features}}.  The combination of the MPO formalism with the cotengra path finder allows us to simulate circuits with very large qubit counts. Specifically, we utilize (i) the advanced cotengra path finder equipped with subtree reconfiguration post-processing, and with a setup similar to Section~\ref{sub:reconf}; (ii) the space efficient MPO representation to evaluate the quantum expectation of the TFIM Hamiltonian (see Section~\ref{sub:mporep}). For the circuit construction, the {\sf split} configuration was used to decompose the parameterized ZZ gates, using SVD to reduce the bond dimension to 2, which further simplifies the tensor network structure to be contracted (see Figure~\ref{fig:split}). Such a two-qubit gate decomposition is implemented as follows. 
\begin{lstlisting}[language=python]
split_conf = {
    "max_singular_values": 2,
    "fixed_choice": 1,
}
# set the SVD decomposition option in circuit level
c = tc.Circuit(n, split=split_conf)
# or set the SVD decomposition option in gate level
c.exp1(i, i + 1, theta=param, unitary=tc.gates._xx_matrix, split=split_conf)
\end{lstlisting}
The ground energy for reference is obtained via two-site DMRG using the Quimb package. 

\bp{Benchmarking}. The parameterized circuit we utilized has $n$ qubits and 7 layers, and contains $21n$ single-qubit gates and $7n$ two-qubit gates. \markedup{We note that our circuit ansatz relies on a succinct tensor network representation for its efficient simulation. In each layer, $n-1$ two-qubit gates are arranged in a cascading ladder layout, i.e. the gates are applied on the (0, 1), (1, 2), (2, 3)...($n-2$, $n-1$) qubit pairs. This ladder layout can spread quantum information and entanglement across the entire chain in only one layer of two-qubit gates, and the resulting state can not readily be simulated by, for instance, taking advantage of a restricted causal light cone as in~\cite{Jobst2022}. }

%Unlike the brickwall layout -- with two-qubit gates layers of -- where the casual light cone for final local operator is restricted and thus efficient simulation on subcircuit is feasible, 
%Therefore, our problem here is intrinsically hard and can not be simulated simply by the light cone idea as presented in \cite{Jobst2022}.

Benchmark results for $n=200, 400, 600$ are summarized in Table~\ref{tab:super}, with times per computational step corresponding to the evaluation of both the energy expectation value and the circuit gradients. \markedup{We find that high-accuracy results are obtained even for large systems without any fine-tuning of the optimizer. This may be explained by the fact that the 1D TFIM has a simple ground state and entanglement structure, and overparametrization (the model has a large number of independent trainable parameters) may also contribute to the accuracy obtained~\cite{Larocca2021a}.}

\begin{table}\centering
\begin{tabular}{c|c|c|c}
   number of qubits  &log2[WRITE]&time for one step& energy accuracy reached\\ 
     \hline 
200 &   27.3    &    5.7   & 99.6\% \\
400 &   29.3    &    11.8  & 99.5\%\\
600 &   31.5    &    18.2   & 99.4\%

\end{tabular}
\caption{Performance benchmarks (running time in seconds) for a large scale VQE task with different qubit counts. \tc results use the TensorFlow backend and run on Nvidia A100 40G GPU. Note that the VQE optimization hyperparameters were not tuned, so the energy accuracy obtained only represents a lower bound on the performance of the current method. Time for one step includes computing both the energy expectation value and the circuit gradients. \markedup{The full optimization requires thousands of iterations, with the Adam gradient descent optimizer used initially, followed by an SGD optimizer in the later stages of the process.}}
\label{tab:super}
\end{table}

\section{Outlook and concluding remarks}

We have introduced \tc, an open-source Python package, designed to meet the requirements of larger and more complex quantum computing simulations. \tc is built on top of, and incorporates, all the main engineering paradigms from modern machine learning libraries, and its flexible and customizable tensor network engine enables high-performance circuit computation.

\bp{Outlook}. We will continue the development of \tc, towards delivering a more efficient and elegant, full-featured and ML-compatible quantum software package. At the top of our priority list are:

\begin{enumerate}
    \item Better tensor network contraction path finders: integrate more advanced algorithms and machine learning techniques for optimal contraction path searching.
    \item Pulse level optimization and quantum control: enable end-to-end differentiable pulse level optimization and optimal quantum control schedules \cite{KHANEJA2005296, 2112.12509}.
    \item Distributed quantum circuit simulation: enable tensor network parallel slicing and distributed computation on multiple hosts.
    \item Approximate circuit simulation based on MPS: introduce TEBD-like algorithms \cite{PhysRevLett.91.147902, PhysRevX.10.041038} to approximately simulate quantum circuits with large size and depth.
    \item More quantum-aware or manifold-aware optimizers: include optimizers such as SPSA~\cite{880982}, rotosolve~\cite{Ostaszewski_2021}, and Riemannian optimizers~\cite{10.21468/SciPostPhys.10.3.079}.
     \item Quantum applications: develop application-level libraries for quantum computing for finance, materials, energy, biology, drug discovery, climate prediction and more.
\end{enumerate}

With the continued rapid progress in quantum computing theory and hardware, our hope is that \tc,  a quantum simulator platform designed for the NISQ era, will play an important role in the academic and commercial progress of this exciting field.

{\bp{Acknowledgements:}} The authors would like to thank our teammates at the Tencent Quantum Laboratory for supporting this project, and Tencent Cloud for providing computing resources. Shi-Xin Zhang would like to thank Rong-Jun Feng, Sai-Nan Huai, Dan-Yu Li, Zi-Xiang Li, Lei Wang, Hao Xie, Shuai Yin and Hao-Kai Zhang for their helpful discussions.

%\bibliographystyle{quantum}
%\bibliography{tc}

\end{document}